%% file: companion.tex
\documentclass[aps,pre,twocolumn,superscriptaddress,longbibliography]{revtex4-2}
\usepackage{amssymb} 
\usepackage{graphicx,color} 
\usepackage{bbm} 
\usepackage{multirow}
\usepackage{amsmath}
\usepackage{mathrsfs} 
\usepackage{hyperref}
\usepackage{bm}
\usepackage[usenames,dvipsnames]{xcolor}
\usepackage{cancel}
\usepackage{color}
\definecolor{darkgreen}{rgb}{0,0.6,0}
\definecolor{darkblue}{rgb}{0,0,0.6}
\definecolor{darkred}{rgb}{0.6,0,0}
\definecolor{darkpurple}{rgb}{0.5,0,0.5}
\usepackage{soul}

\hypersetup{
bookmarksopen=true,
pdftitle="",
pdfauthor="", 
pdftoolbar=false, 
pdfstartview={FitH},		
pdfmenubar=true,			
pdfhighlight=/O,			
colorlinks=true,			
urlcolor=darkblue,
citecolor=darkblue,		
linkcolor=darkblue}

\usepackage{tikz}
 
\input{gp_macros}
\input{macros_diagrams}

\usetikzlibrary{patterns}
\usetikzlibrary{decorations.pathmorphing}
\usetikzlibrary{decorations.markings}
 \newcommand{\e}[1]{e^{#1}}

  \usetikzlibrary{snakes}

\usepackage{CJK}

\newcommand{\tildephi}{\widetilde\phi}
\newcommand{\tildepsi}{\widetilde{\psi}}

\newcommand{\Prob}[1]{P_{\hspace{-2pt}{\scalebox{.8}{$\scriptscriptstyle #1$}}}}
\newcommand{\Ppp}{\Prob{++}}
\newcommand{\Ppm}{\Prob{+-}}
\newcommand{\Pmp}{\Prob{-+}}

\newcommand{\Pppj}[1]{P_{\hspace{-2pt}{\scalebox{.8}{$\scriptscriptstyle ++$}},#1}}
\newcommand{\Pmpj}[1]{P_{\hspace{-2pt}{\scalebox{.8}{$\scriptscriptstyle -+$}},#1}}

\newcommand{\Pe}{\mathsf{Pe}}
\newcommand{\xibar}{\bar{\xi}}
\newcommand{\nubar}{\bar{\nu}}
\newcommand{\gammabar}{\bar{\gamma}}

\DeclareMathOperator{\realpart}{Re}
\DeclareMathOperator{\imaginarypart}{Im}
\DeclareMathOperator{\sign}{sign}
\newcommand{\epr}{\dot{\mathcal{S}}}

\newcommand{\Resm}[2]{\operatorname{Res}_{#1}#2}

\newcommand{\zilname}{\begin{CJK*}{UTF8}{gbsn}张子洛\end{CJK*}}

\allowdisplaybreaks

\newcommand{\titleText}{Microscopic theory of soft run-and-tumble particles}
\title{\titleText}

\CJKnospace
\begin{document}

\title{\titleText}
\author{Rosalba Garcia-Millan}
\affiliation{Department of Mathematics, King's College London, Strand, London WC2R 2LS, United Kingdom}
\affiliation{DAMTP, Centre for Mathematical Sciences, University of Cambridge, Cambridge CB3 0WA, United Kingdom}
\affiliation{St John's College, University of Cambridge, Cambridge CB2 1TP, United Kingdom}
\author{Ziluo Zhang (\zilname)}
\affiliation{Department of Mathematics and Centre of Complexity Science, Imperial College London, London SW7 2AZ, United Kingdom}
\author{Luca Cocconi}
\affiliation{Department of Mathematics and Centre of Complexity Science, Imperial College London, London SW7 2AZ, United Kingdom}
\author{Marius Bothe}
\affiliation{Department of Mathematics and Centre of Complexity Science, Imperial College London, London SW7 2AZ, United Kingdom}
\author{Letian Chen}
\affiliation{Department of Mathematics and Centre of Complexity Science, Imperial College London, London SW7 2AZ, United Kingdom}
\author{Zigan Zhen}
\affiliation{Department of Mathematics and Centre of Complexity Science, Imperial College London, London SW7 2AZ, United Kingdom}
\author{Gunnar Pruessner}
\email{g.pruessner@imperial.ac.uk}
\affiliation{Department of Mathematics and Centre of Complexity Science, Imperial College London, London SW7 2AZ, United Kingdom}

\date{\today}

\begin{abstract}
Soft, repulsive run-and-tumble particles display emergent effective interactions
as they appear to stick to each other in spite of the absence of attractive forces.
This effective attraction emerges at strong enough repulsion and large 
self-propulsion.
Complementing a companion paper that characterises effective attraction between two soft run-and-tumble particles
\cite{letter}, here we provide a thorough derivation of our microscopic theory, 
which is an exact representation of the particle dynamics.
We report the systematic calculation of the effective interaction vertices iteratively,
in a perturbation expansion about the interaction couplings, by adding, order by order, loop corrections.
We use the effective interaction vertices to calculate the
two-point correlation function, fully characterising
the stationary state.
Other observables, such as the structure factor, overlap probability and entropy production rate are 
calculated as well.
\end{abstract}

\maketitle

\section{Introduction}

Active particles are out of thermodynamic equilibrium by constantly transforming a local fuel into mechanical work.
A notable phenomenon in active particle systems is the emergence of Motility-Induced Phase Separation
(MIPS), where self-propelled particles that interact repulsively tend to form clusters, separating in space
into ``dilute" and ``liquid" phases
\cite{FilyMarchetti:2012,
CatesTailleur:2015,
DigregorioETAL:2018,
CatesNardini:2025}.
This \emph{stickiness} between otherwise repulsive particles
can be understood as 
the emergence of effective attraction between them
\cite{SlowmanETAL:2016,
GuillinHahnMichel:2025,
DasDharKundu:2020,
HahnGuillinMichel:2025,
MetsonEvansBlythe:2023,
preprintHahnGuillinMichel:2025}.
Computer simulations show that this effect also manifests
between two soft, self-propelled particles in a one-dimensional
space.
In this paper, we provide the analytical framework of the findings in \cite{letter}.
Our results establish the analytic relation between microscopic dynamics and emergence of
effective interactions between two soft repulsive self-propelled particles in a continuous periodic domain.

The contents of this paper are organised as follows.
We define the model of $N$ soft run-and-tumble particles (RTPs) on
a ring interacting via a repulsive Yukawa potential
in \sref{model}.
Using the equation of motion, given by an overdamped Langevin equation,
we derive the Fokker-Planck equation that governs the evolution of the joint particle number density
in \sref{DPFT}. From the Fokker-Planck equation we derive
the Doi-Peliti \cite{Doi:1976, Doi:Mar1976, Peliti:1985, Cardy:2006}
action functional following the framework in Ref.~\cite{PruessnerGarcia-Millan:2025}.
The action has a Gaussian part that provides the (Dyson-summed) propagators
and a non-bilinear part that provides the perturbative couplings or bare interaction vertices.

Quantifying effective interactions in the stationary state reduces to extracting the relevant properties of
the pair-correlation function,
or its counterpart in Fourier space, which is the static structure factor \cite{FilyMarchetti:2012,HansenMcDonald:2006}.
Effective interactions are well characterised by the lowest Fourier mode of the structure factor,
which is, in turn, intimately related to the compressibility and the mean-square distance 
\cite{letter,HansenMcDonald:2006}.
Characterising the stationary two-point correlation function analytically requires the 
renormalisation of the bare interaction vertices, which we calculate in a perturbation expansion about
the interaction coupling. Each order in the expansion is given by a new loop correction
to the two-point vertices that accounts for self-propulsion, tumbling, diffusion and bare interaction.
The sum of all terms in the perturbation expansion gives the effective interaction vertices,
which are the key mathematical object needed to calculate the stationary two-point correlation functions.
In \sref{eff_vertices}, we devise an iterative method to calculate the effective interaction vertices order
by order in the interaction coupling analytically.

In \sref{observables} we apply the effective interaction vertices to calculate the following observables:
the static structure factor, two-point correlation functions, the overlap probability
and the entropy production rate.
These observables are complementary to the compressibility discussed in \cite{letter}, 
consistently reporting on the emergence of effective attraction between active particles.
We conclude in \sref{conclusion} with a discussion and an outlook.

\section{Soft run-and-tumble particles}
\seclabel{model}
We consider the model studied in \cite{letter} of $N$ interacting RTPs on a periodic one-dimensional domain
at positions $x_i\in[-L/2,L/2)$.
The coupled overdamped Langevin equations of motion are
\begin{align}
\elabel{oLEq}
\dot{x}_i =	 \drift\sigma_i(t) - \sum_{j=1}^N\pairPot'(x_i-x_j) + \sqrt{2\diff}\eta_i(t) \ ,
\end{align}
where $\drift$ is the in-built particle self-propulsion velocity,
the orientation $\sigma_i$ is a telegraphic noise that switches
between values $\pm1$ with Poissonian rate $\gamma$,
$\eta_i$ is a thermal noise modelled by a
unit Gaussian white noise with $\ave{\eta_i(t)}=0$ and
$\ave{\eta_i(t)\eta_j(t)}=\delta_{i,j}\delta(t-t')$ and
$\diff$ is the diffusion constant.
We model \emph{soft} particle repulsion at distance $x=|x_i-x_j|$ 
with interaction forces derived from a Yukawa potential,
\begin{equation}
    \pairPotInf(x) =  \frac{\nu}{2\xi} \exp{-|x|/\xi}
    \elabel{pair_potentialInf}
\end{equation}
with interaction length $\xi$ and coupling $\nu$.
Since particles move in a bounded periodic domain, 
 the total interaction force experienced by a particle is the sum
of forces at distance $x$ \textit{modulo} the system size $L$.
The Yukawa potential on a periodic domain $x\in[-L/2,L/2)$ is then
\begin{align}
    \pairPot(x) = & \sum_{m=-\infty}^\infty \pairPotInf(x+mL) \nonumber\\
    = & \frac{\nu \cosh((|x|-L/2)/\xi)}{2\xi \sinh(L/(2\xi))} \ ,
    \elabel{pair_potentialRing}
\end{align}
which approaches $\pairPotInf(x)$ as $\xi\ll L$.
Since interaction forces are finite for all $x$, the two particles can overlap and overcome the potential barrier at $x=0$. 
We recover excluded-volume interactions in the limit $\xi\to0$, though this scenario is not
considered in the present work.

\section{Doi-Peliti action functional}
\seclabel{DPFT}
The stochastic process governed by \Eref{oLEq} is equivalently described by the Fokker-Planck equation for the
\emph{joint particle number density} of the vector of positions $\xvec=(x_1,\ldots,x_N)$ and internal states
$\sigmavec=(\sigma_1,\ldots,\sigma_N)$
 \cite{ZhangPruessner:2022,PruessnerGarcia-Millan:2025,ZhangGarcia-Millan:2023},
\begin{align}
\elabel{FPE}
\partial_t\rho(\xvec;\sigmavec;t) = &
\sum_{i=1}^N \Big\{\left(\diff \partial_{x_i}^2 - \drift\sigma_i \partial_{x_i} - 2\gamma\right) \rho(\xvec;\sigmavec;t) 
\nonumber\\
&
+{\gamma}\sum_{\sigma_i} \rho(\xvec;\sigmavec;t)\nonumber\\
&
+\partial_{x_i}\Big(
\sum_{\substack{{j=1}\\{j\neq i}}}^N (\partial_{x_i}\pairPot(x_i-x_j)) \rho(\xvec;\sigmavec;t)
\Big)
\Big\}\ .
\end{align}

The Doi-Peliti action functional $\MA$ follows immediately from the Fokker-Planck \Eref{FPE} 
\cite{PruessnerGarcia-Millan:2025}.
The action does not contain more or less information than what is provided by the Fokker-Planck equation
and the field theory is exact in the sense that it does not entail any coarse graining or approximation.
Observables are normally expressed in an order-by-order expansion of the non-linear couplings using the
language of diagrams and without losing the notion of particulate degrees of freedom \cite{BotheETAL:2023}.
Since the internal state of each particle is discrete, we consider for each state different species of the field:
we use the fields $\phi$, $\tildephi$ for right-moving particles ($\sigma_i=1$), and $\psi$, $\tildepsi$ for left-moving particles ($\sigma_i=-1$), where $\phi$ and  $\psi$ are annihilation fields, and $\tildephi$ and $\tildepsi$ are Doi-shifted creation fields \cite{ZhangPruessner:2022,Garcia-MillanPruessner:2021}.
Separating the free particle dynamics from particle interactions, the Doi-Peliti action functional $\MA = \MA_0 + \MA_1 $ can be split into two parts \cite{PruessnerGarcia-Millan:2025}: 
the Gaussian part (or bilinear part) of the action,
\begin{align}
\elabel{A0}
\MA_0=- \int \dint{x}\dint{t}
 \{ & \tildephi(\partial_t +\drift\partial_x-\diff\partial_x^2 )\phi \nonumber\\
&+\tildepsi(\partial_t -\drift\partial_x-\diff\partial_x^2 )\psi  \nonumber\\
&+\gamma(\tilde \phi-\tilde \psi)(\phi-\psi) \}
\end{align}
and the perturbative part of the action,
\begin{widetext}
\begin{align}
\MA_1=
-&\int \dint{x} \dint{y} \dint{t}
\left(
\left(1+\tildephi(y,t)\right)\phi(y,t)
+
\left(1+\tildepsi(y,t)\right)\psi(y,t)
\right)
\partial_x\pairPot(x-y)
\left(
\phi(x,t)\partial_x\tildephi(x,t)
+
\psi(x,t)\partial_x\tildepsi(x,t)
\right) \ .
\elabel{A1}
\end{align}
\end{widetext}
In this theory, the action
does not depend on the number $N$ of particles in the system.
Instead, the number of particles is fixed by the initialisation at a later stage in the derivation.
We adopt the following sign convention for the action in the path integral
when calculating an observable $\OC$,
\begin{equation}
\ave{\OC} = \int\mathcal{D}[\phi,\tildephi,\psi,\tildepsi] \OC \exp{\MA[\phi,\tildephi,\psi,\tildepsi]}  \ .
\end{equation}

\subsection{Fourier convention}
We use the following Fourier transforms 
\begin{subequations}
\begin{align}
    \phi(x,t) & = \frac{1}{L} \sum_{j=-\infty}^{\infty}
    \int_{-\infty}^\infty \dintbar{\omega} 
    \exp{-\imag\omega t} \exp{\imag k_j x}
    \phi_{j}(\omega) \ ,\\
    \phi_{j}(\omega) & = \int_{-L/2}^{L/2} \dint{x} \int_{-\infty}^\infty \dint{t} 
    \exp{\imag\omega t} \exp{-\imag k_j x}
    \phi(x,t) \ ,
    \elabel{Fourier_convention}
\end{align}
\end{subequations}
and correspondingly for $\tildephi$, $\psi$ and $\tildepsi$,
with wavenumber $k_j=2\pi j/L$ and frequency $\omega$.
These transforms obey the orthogonality relations
\begin{subequations}
\begin{align}
 \int_{-\infty}^\infty \dint{t} 
    \exp{\imag\omega t} 
    = & \deltabar(\omega) \ , \\
 \int_{-\infty}^\infty \dintbar{\omega} 
    \exp{-\imag\omega t} 
    = & \delta(t) \ , 
    \\
    \int_{-L/2}^{L/2} \dint{x}  \exp{-\imag k_j x}
    = & L \delta_{j,0} \ , \elabel{K_delta}\\
    \frac{1}{L} \sum_{j=-\infty}^\infty \exp{\imag k_j x} = &
    \sum_{m=-\infty}^\infty \delta(x+mL) \ , \elabel{F_sum}
\end{align}
\end{subequations}
where we have used the notation $\dbar{\omega}=\dint{\omega}/(2\pi)$,
the Dirac $\delta$-function $\deltabar(\omega)=2\pi\delta(\omega)$,
and the Kronecker $\delta$-function $\delta_{j,0}$.
For easier book keeping, we recognise that any Kronecker $\delta$-function
that arises from orthogonality is preceded by a factor $L$, \Eref{K_delta},
and
sums over Fourier modes
are preceded by a factor $1/L$, \Eref{F_sum}, dimensionally consistent with Fourier integrals.
The (bare) interaction potential in \Eref{pair_potentialRing} has the convenient Fourier transform 
\begin{equation}
\elabel{barePotF}
    \pairPot_j = \int_{-L/2}^{L/2} \dint{x} \exp{-\imag k_j x} \pairPot(x) =
    \frac{\nu\xi^{-2}}{k_j^2+\xi^{-2}}
\ .
\end{equation}

\subsection{Propagators and interaction vertices}
\seclabel{props_n_vertices}
The propagators and interaction vertices are obtained by
transforming the action functional $\MA$ to Fourier space.
To ease notation we introduce the shorthands
\begin{subequations}
\elabel{prop_conventions}
\begin{align}
    \elabel{prop_conventions_G}
    G(k,\omega)=&-\imag\omega + \diff k^2 - \imag \drift k + \gamma  \\
    \elabel{prop_conventions_E}
    E(k,\omega)=&-\imag\omega + \diff k^2 + \imag \drift k + \gamma  \\
    H=&\gamma\\
    \elabel{prop_conventions_D}
    D(k,\omega)=&(GE-H^2)^{-1} \ ,
\end{align}
\end{subequations}
which allows us to write the bare propagators of the harmonic part of the action as
\begin{subequations}
\elabel{bare_props}
\begin{align}
\tikz[baseline=-2.5pt]{
\draw[Aactivity] (0,0) -- (1,0);
} \corresponds & \langle{\phi_j(\omega)\tildephi_{j'}(\omega')}\rangle_0 
\nonumber\\
= &
L \delta_{j+j',0} \deltabar(\omega+\omega') \ D(k_j,\omega) G(k_j,\omega)\\
\tikz[baseline=-2.5pt]{
\draw[Aactivity] (0,0) -- (0.5,0);
\draw[tAsubstrate] (0.5,0) -- (1,0);
} 
\corresponds & \langle{\phi_j(\omega)\tildepsi_{j'}(\omega')}\rangle_0 
\nonumber\\
= &
 L \delta_{j+j',0} \deltabar(\omega+\omega') \ D(k_j,\omega) H\\
\tikz[baseline=-2.5pt]{
\draw[tAsubstrate] (0,0) -- (0.5,0);
\draw[Aactivity] (0.5,0) -- (1,0);
} 
\corresponds & \langle{\psi_j(\omega)\tildephi_{j'}(\omega')}\rangle_0 
\nonumber\\
= &
 L \delta_{j+j',0} \deltabar(\omega+\omega') \ D(k_j,\omega) H\\
\tikz[baseline=-2.5pt]{
\draw[tAsubstrate] (0,0) -- (1,0);
} \corresponds & \langle{\psi_j(\omega)\tildepsi_{j'}(\omega')}\rangle_0 
\nonumber\\
= &
 L \delta_{j+j',0} \deltabar(\omega+\omega') \ D(k_j,\omega) E(k_j,\omega) \ .
\end{align}
\end{subequations}

The perturbative part of the action in \Eref{A1} has a total of eight interaction vertices: four three-point vertices and another four four-point vertices. Of all the vertices, only the four-point vertices enter in the observables that we 
consider here for the two-particle case,
\begin{subequations}
\elabel{vertices}
\begin{align}
\elabel{vertex_bare_homo_straight}
 \tikz[baseline=-2.5pt]{
    \draw[Aactivity] (-0.4,0.4) node[left] {\footnotesize $1$} -- (0,0.3) -- (0.4,0.4) node[right] {\footnotesize $2$};
    \draw[draw=none] (-0.4,0.4) -- 
      node[pos=0.4,sloped] {\tikz{\draw (0,-1.2mm) -- (0,1.2mm);}} (0,0.3) ;
    \draw[DasheDpotential] (0,0.3) -- (0,-0.3);
    \draw[Aactivity] (-0.4,-0.4) node[left] {\footnotesize $3$} -- (0,-0.3) -- (0.4,-0.4) node[right] {\footnotesize $4$};
    }
   \corresponds &
   L\delta_{j_1+j_2+j_3+j_4,0}\deltabar(\omega_1+\omega_2+\omega_3+\omega_4)
    \nonumber\\&
   \pairPot_{j_3+j_4}
   k_{j_1} k_{j_3+j_4}  
   \\
\elabel{vertex_bare_homo_wriggly}
     \tikz[baseline=-2.5pt]{
    \draw[tAsubstrate] (-0.4,0.4) node[left] {\footnotesize $1$} -- (0,0.3) -- (0.4,0.4) node[right] {\footnotesize $2$};
    \draw[draw=none] (-0.4,0.4) -- 
      node[pos=0.4,sloped] {\tikz{\draw (0,-1.2mm) -- (0,1.2mm);}} (0,0.3) ;
    \draw[DasheDpotential] (0,0.3) -- (0,-0.3);
    \draw[tAsubstrate] (-0.4,-0.4) node[left] {\footnotesize $3$} -- (0,-0.3) -- (0.4,-0.4) node[right] {\footnotesize $4$};
    } 
   \corresponds &
   L\delta_{j_1+j_2+j_3+j_4,0}\deltabar(\omega_1+\omega_2+\omega_3+\omega_4)
    \nonumber\\&
   \pairPot_{j_3+j_4} 
   k_{j_1} k_{j_3+j_4}  
   \\
\elabel{vertex_bare_hetero_straight_wriggly}
    \tikz[baseline=-2.5pt]{
    \draw[Aactivity] (-0.4,0.4) node[left] {\footnotesize $1$} -- (0,0.3) -- (0.4,0.4) node[right] {\footnotesize $2$};
    \draw[draw=none] (-0.4,0.4) -- 
      node[pos=0.4,sloped] {\tikz{\draw (0,-1.2mm) -- (0,1.2mm);}} (0,0.3) ;
    \draw[DasheDpotential] (0,0.3) -- (0,-0.3);
    \draw[tAsubstrate] (-0.4,-0.4) node[left] {\footnotesize $3$} -- (0,-0.3) -- (0.4,-0.4) node[right] {\footnotesize $4$};
    }
   \corresponds &
   L\delta_{j_1+j_2+j_3+j_4,0}\deltabar(\omega_1+\omega_2+\omega_3+\omega_4)
    \nonumber\\&
   \pairPot_{j_3+j_4} 
   k_{j_1} k_{j_3+j_4}  
   \\
\elabel{vertex_bare_hetero_wriggly_straight}
    \tikz[baseline=-2.5pt]{
    \draw[tAsubstrate] (-0.4,0.4) node[left] {\footnotesize $1$} -- (0,0.3) -- (0.4,0.4) node[right] {\footnotesize $2$};
    \draw[draw=none] (-0.4,0.4) -- 
      node[pos=0.4,sloped] {\tikz{\draw (0,-1.2mm) -- (0,1.2mm);}} (0,0.3) ;
    \draw[DasheDpotential] (0,0.3) -- (0,-0.3);
    \draw[Aactivity] (-0.4,-0.4) node[left] {\footnotesize $3$} -- (0,-0.3) -- (0.4,-0.4) node[right] {\footnotesize $4$};
    }
   \corresponds &
   L\delta_{j_1+j_2+j_3+j_4,0}\deltabar(\omega_1+\omega_2+\omega_3+\omega_4)
    \nonumber\\&
   \pairPot_{j_3+j_4} 
   k_{j_1} k_{j_3+j_4}  
\end{align}
\end{subequations}
These four interaction vertices originate in the four combinations of forces between two species of particles:
\eref{vertex_bare_homo_straight} captures the force exerted on a right-moving particle by another right-moving particle;
\eref{vertex_bare_homo_wriggly}, the force by a left-moving particle on a left-moving particle;
\eref{vertex_bare_hetero_straight_wriggly}, the force by a left-moving particle on a right-moving particle; and
\eref{vertex_bare_hetero_wriggly_straight}, the force by a right-moving particle on a left-moving particle.

\section{Effective interaction vertices}
\seclabel{eff_vertices}
The main observable considered in the present work is the stationary two-point correlation function.
Calculating it analytically requires a renormalisation of the interaction vertices, incorporating
all loop corrections to the four-point interaction vertices in \eref{vertices}.
We study the two-particle case $N=2$, 
where the derivation does not involve three-point interaction vertices \cite{ZhangGarcia-Millan:2023},
leaving the many-particle case for future work.

We therefore formally initialise the present system with two creator fields at time $t_0$ at $x_{01}$ and $x_{02}$, so that any observable $\OC$ is preceded by, say $\phi^\dagger(x_{02},t_0)\phi^\dagger(x_{01},t_0)$. In the case of the two-point correlations studied in the following, the only non-vanishing contributions are due to the Doi-shifted fields $\tildephi(x_{02},t_0)\tildephi(x_{01},t_0)$. To study the steady state, this initialisation ought to take place at $t_0\to-\infty$. Taking this limit by Fourier-transforming observables of the form $\ave{\OC \tildephi_{j_{02}}(\omega_{02})\tildephi_{j_{01}(\omega_{01})}}$ or by using the final value theorem (or terminal-value theorem) \cite{Schiff:1999} reveals that the limit solely affects the incoming legs, effectively reducing each such leg to a uniform $1/(2L)$ from the joint
probability of the particle position and internal state at stationarity,
\begin{align}
& \lim_{t_0\to -\infty}  \langle\phi(x,t)\tildephi(x_0,t_0)\rangle = 
\lim_{t_0\to -\infty}\langle\psi(x,t)\tildepsi(x_0,t_0)\rangle = 
\nonumber\\  & =
\lim_{t_0\to -\infty}\langle\phi(x,t)\tildepsi(x_0,t_0)\rangle = 
\lim_{t_0\to -\infty}\langle\psi(x,t)\tildephi(x_0,t_0)\rangle  
\nonumber\\  & =
\frac{1}{2L} \ .
\elabel{steady_state_in_legs}
\end{align}
In \Eref{example_Phi} this results in a factor $1/(2L)^2$
from incoming legs  times a symmetry factor of $2$ 
for the two possible arrangements of incoming legs.
The uniform distribution in realspace transforms into a $\delta$-function in Fourier space, producing
$L\delta_{j_2,0}\deltabar(\omega_2)$ and
$L\delta_{j_4,0}\deltabar(\omega_4)$
for each incoming leg respectively, fixing the incoming wavenumber and frequency to zero
in the vertices \eref{vertices} and leaving an overall prefactor of $1/2$.
Diagrammatically, the initialisation in the far past, $t_0\to -\infty$, thus amounts to replacing the incoming legs
by a constant pre-factor $1/(2L^2)$ and a removal of these legs in their entirety
\cite{ZhenPruessner:2022,ZhangPruessner:2022,ZhangGarcia-Millan:2022},
beyond a common amputation \cite{LeBellac:1991}.

We define the following effective interaction vertices, 
which include all contributions from diagrams with no incoming legs and matching amputated outgoing legs,
\begin{subequations}
\elabel{PotsAsSums}
\begin{align}
    \HomoPot(k_j) &\corresponds 
    \tikz[baseline=-2.5pt]{
    \draw[pattern=north west lines] (0,0) circle (0.7cm);
    \draw[fill=white] (0,0) node[circle,fill=white,inner sep=2pt,opacity=0.9]  {$\HomoPot(k_j)$};
    \draw[DAtAactivity] (150:0.7cm) -- (150:1.15cm) node[left] {$k_j$};
    \draw[draw=none] (150:0.7cm) -- node[pos=0.4,above=0.1cm] {$-k_j$} (150:1.05cm) ;
    \draw[Aactivity] (210:0.7cm) -- (210:1.05cm) node[left] {$\ $}; 
}    
    \ , &
    \tHomoPot(k_j) &\corresponds
    \tikz[baseline=-2.5pt]{
    \draw[pattern=north west lines] (0,0) circle (0.7cm);
    \draw[fill=white] (0,0) node[circle,fill=white,inner sep=2pt,opacity=0.9]  {$\tHomoPot(k_j)$}; 
    \draw[tAsubstrate] (150:0.7cm) -- (150:1.05cm) node[pos=0.4,sloped] {\tikz{\draw (0,-1.2mm) -- (0,1.2mm);}};
    \draw[tAsubstrate] (150:0.7cm) -- node[pos=0.4,above=0.1cm] {$-k_j$} (150:1.05cm) node[left] {$k_j$}; 
    \draw[tAsubstrate] (210:0.7cm) -- (210:1.05cm); 
    } \ ,
\\
    \HeteroPot(k_j) &\corresponds
    \tikz[baseline=-2.5pt]{
    \draw[pattern=north west lines] (0,0) circle (0.7cm);
    \draw[fill=white] (0,0) node[circle,fill=white,inner sep=2pt,opacity=0.9]  {$\HeteroPot(k_j)$};
    \draw[tAsubstrate] (150:0.7cm) -- (150:1.05cm) node[pos=0.4,sloped] {\tikz{\draw (0,-1.2mm) -- (0,1.2mm);}} node[left] {$k_j$};
    \draw[draw=none] (150:0.7cm) -- node[pos=0.4,above=0.1cm] {$-k_j$} (150:1.05cm) ;
    \draw[Aactivity] (210:0.7cm) -- (210:1.05cm); 
    }
    \ , &
\tHeteroPot(k_j) &\corresponds
    \tikz[baseline=-2.5pt]{
    \draw[pattern=north west lines] (0,0) circle (0.7cm);
    \draw[fill=white] (0,0) node[circle,fill=white,inner sep=2pt,opacity=0.9]  {$\tHeteroPot(k_j)$};
    \draw[Aactivity] (150:0.7cm) -- (150:1.05cm) node[pos=0.4,sloped] {\tikz{\draw (0,-1.2mm) -- (0,1.2mm);}} node[left] {$k_j$};
    \draw[draw=none] (150:0.7cm) -- node[pos=0.4,above=0.1cm] {$-k_j$} (150:1.05cm) ;
    \draw[tAsubstrate] (210:0.7cm) -- (210:1.05cm); 
    } \ .
\end{align}
\end{subequations}
These four effective vertices encapsulate the aggregate effect of particle interactions and can
be used to calculate observables in the stationary state.

A perturbation expansion in the repulsion $\nu$ shows that the correction terms of each
effective interaction vertex at order $\nu^n$ are interdependent.
For instance, to one loop the contributions to $\HomoPot$ are
\begin{align}
\elabel{example_Phi}
    \HomoPot(k_j) \corresponds 
    &
\tikz[baseline=-2.5pt]{
    \draw[AactivityDA] (-0.4,0.4) -- (0,0.3);
    \draw[DasheDpotential] (0,0.3) -- (0,-0.3);
    \draw[Aactivity] (-0.4,-0.4) -- (0,-0.3);
    }
    +
    \LoopDiagramfi{Aactivity}{}{Aactivity}{}{}{Aactivity}{Aactivity}{draw=none}{}
    +
    \LoopDiagramreverse{Aactivity}{}{Aactivity}{draw=none}{}{Aactivity}{Aactivity}{}{}
    +
    \LoopDiagramfi{Aactivity}{}{tAsubstrate}{}{}{Aactivity}{Aactivity}{draw=none}{}
   \nonumber\\ &
    +
    \LoopDiagramreverse{Aactivity}{}{tAsubstrate}{draw=none}{}{Aactivity}{Aactivity}{}{}
    +
    \LoopDiagramfi{Aactivity}{}{Aactivity}{}{}{Aactivity}{tAsubstrate}{draw=none}{}
    +
    \LoopDiagramreverse{Aactivity}{}{Aactivity}{draw=none}{}{Aactivity}{tAsubstrate}{}{}
   \nonumber\\ &
    +
    \LoopDiagramfi{Aactivity}{}{tAsubstrate}{}{}{Aactivity}{tAsubstrate}{draw=none}{}
    +
    \LoopDiagramreverse{Aactivity}{}{tAsubstrate}{draw=none}{}{Aactivity}{tAsubstrate}{}{}
    +\OC(\nu^3)
\end{align}

For easier accounting we write the perturbative expansion of $\HomoPot$ in $\nu$ as
\begin{equation}
\HomoPot = \sum_{n\geq1}\HomoPot_n
\end{equation}
with $\HomoPot_n\propto\nu^n$.
Similarly, we define the $n$-th order in $\nu$ for the other three effective vertices,
$\HeteroPot_n$, $\tHomoPot_n$ and $\tHeteroPot_n$.

Using the explicit form of the bare interaction potential \eref{barePotF} in \Eref{vertices}
with incoming legs in the steady state, \Eref{steady_state_in_legs}, and amputated outgoing legs, the
first-order term of the effective interaction vertices are all identical
\begin{subequations}
\elabel{pots1}
\begin{align}
\elabel{HomoPot_1}
    \HomoPot_1(k_j)
    =& \frac{1}{2L^2}
    \frac{\nu\xi^{-2}}{k_j^2+\xi^{-2}} (-k_j)k_j
    \corresponds
    \tikz[baseline=-2.5pt]{
    \draw[Aactivity] (-0.4,0.4) -- (0,0.3);
    \draw[draw=none] (-0.4,0.4) -- 
      node[pos=0.4,sloped] {\tikz{\draw (0,-1.2mm) -- (0,1.2mm);}}
      node[pos=0.4,above=0.1cm] {$-k_j$} (0,0.3) ;
    \draw[DasheDpotential] (0,0.3) -- (0,-0.3);
    \draw[Aactivity] (-0.4,-0.4) -- (0,-0.3);
    } \\
\elabel{HeteroPot_1}
    \HeteroPot_1(k_j)
    =& \frac{1}{2L^2}
    \frac{\nu\xi^{-2}}{k_j^2+\xi^{-2}} (-k_j)k_j
    \corresponds
    \tikz[baseline=-2.5pt]{
    \draw[tAsubstrate] (-0.4,0.4) -- (0,0.3);
    \draw[draw=none] (-0.4,0.4) -- 
      node[pos=0.4,sloped] {\tikz{\draw (0,-1.2mm) -- (0,1.2mm);}}
      node[pos=0.4,above=0.1cm] {$-k_j$} (0,0.3) ;
    \draw[DasheDpotential] (0,0.3) -- (0,-0.3);
    \draw[Aactivity] (-0.4,-0.4) -- (0,-0.3);
    } \\
    \tHomoPot_1(k_j)
    =& \frac{1}{2L^2}
    \frac{\nu\xi^{-2}}{k_j^2+\xi^{-2}} (-k_j)k_j
    \corresponds
    \tikz[baseline=-2.5pt]{
    \draw[tAsubstrate] (-0.4,0.4) -- (0,0.3);
    \draw[draw=none] (-0.4,0.4) -- 
      node[pos=0.4,sloped] {\tikz{\draw (0,-1.2mm) -- (0,1.2mm);}}
      node[pos=0.4,above=0.1cm] {$-k_j$} (0,0.3) ;
    \draw[DasheDpotential] (0,0.3) -- (0,-0.3);
    \draw[tAsubstrate] (-0.4,-0.4) -- (0,-0.3);
    } \\
    \tHeteroPot_1(k_j)
    =& \frac{1}{2L^2}
    \frac{\nu\xi^{-2}}{k_j^2+\xi^{-2}} (-k_j)k_j
    \corresponds
    \tikz[baseline=-2.5pt]{
    \draw[Aactivity] (-0.4,0.4) -- (0,0.3);
    \draw[draw=none] (-0.4,0.4) -- 
      node[pos=0.4,sloped] {\tikz{\draw (0,-1.2mm) -- (0,1.2mm);}}
      node[pos=0.4,above=0.1cm] {$-k_j$} (0,0.3) ;
    \draw[DasheDpotential] (0,0.3) -- (0,-0.3);
    \draw[tAsubstrate] (-0.4,-0.4) -- (0,-0.3);
    } \ ,
\end{align}
\end{subequations}
which are defined to {not} contain the $L\delta$ and $\deltabar$ functions due to translational invariance
in space and time. 
The origin of
the overall factor of $1/(2L^2)$ in \Eref{pots1} is discussed after \Eref{steady_state_in_legs}.

\subsection{Iteration}
Calculating the second-order corrections $\HomoPot_2$, $\HeteroPot_2$, $\tHomoPot_2$ and $\tHeteroPot_2$
is a matter of summing over the loop corrections generated by attaching 
the vertex with the correct (amputated) outgoing legs in \Erefs{vertices}
to each first-order correction in \Erefs{pots1} in every orientation. 
This iteration, which increases the number of terms by a factor 8, generalises to any order $n$.
For instance, the correction $\HomoPot_{n+1}$ is generated from 
$\HomoPot_n$, $\HeteroPot_n$, $\tHomoPot_n$ and $\tHeteroPot_n$ diagrammatically as follows,
\begin{align}
\elabel{HomoPot_recurrence_diag}
     \HomoPot_{n+1}(k_j)  &\corresponds 
\renewcommand{\LoopDiagramPot}{\HomoPot_n(k_i)}
\LoopDiagram{Aactivity}{}{Aactivity}{}{-k_i}{Aactivity}{Aactivity}{draw=none}{}
\ +
\renewcommand{\LoopDiagramPot}{\HomoPot_n(-k_i)}
\LoopDiagram{Aactivity}{}{Aactivity}{draw=none}{}{Aactivity}{Aactivity}{}{k_i}
\nonumber \\ &
\ +
\renewcommand{\LoopDiagramPot}{\HeteroPot_n(k_i)}
\LoopDiagram{Aactivity}{}{tAsubstrate}{}{-k_i}{Aactivity}{Aactivity}{draw=none}{}
\ +
\renewcommand{\LoopDiagramPot}{\tHeteroPot_n(-k_i)}
\LoopDiagram{Aactivity}{}{tAsubstrate}{draw=none}{}{Aactivity}{Aactivity}{}{k_i} \nonumber\\
\nonumber&+
\renewcommand{\LoopDiagramPot}{\tHeteroPot_n(k_i)}
\LoopDiagram{Aactivity}{}{Aactivity}{}{-k_i}{Aactivity}{tAsubstrate}{draw=none}{}
\ +
\renewcommand{\LoopDiagramPot}{\HeteroPot_n(-k_i)}
\LoopDiagram{Aactivity}{}{Aactivity}{draw=none}{}{Aactivity}{tAsubstrate}{}{k_i}
\nonumber \\ &
\ +
\renewcommand{\LoopDiagramPot}{\tHomoPot_n(k_i)}
\LoopDiagram{Aactivity}{}{tAsubstrate}{}{-k_i}{Aactivity}{tAsubstrate}{draw=none}{}
\ +
\renewcommand{\LoopDiagramPot}{\tHomoPot_n(-k_i)}
\LoopDiagram{Aactivity}{}{tAsubstrate}{draw=none}{}{Aactivity}{tAsubstrate}{}{k_i} \ .
\end{align}

Using the notation $G=G(k_i,\omega)$ and $G'=G(-k_i,-\omega)$, etc.,
for the terms defined in \Erefs{prop_conventions}
and, similarly $\HomoPot=\HomoPot(k_i)$ and $\HomoPot'=\HomoPot(-k_i)$ for the effective interaction vertices,
the iterative step to generate the $n+1$-th order is, 
\begin{subequations}
\elabel{PotsRecursion}
\begin{align}
\elabel{HomoPot_recurrence}
\HomoPot_{n+1}(k_j)
=&(-k_j)
 \frac{1}{L}
    \sum_{
    \substack{i\in\Zset\\
    i\ne0}}
    \int\dintbar{\omega'} 
 \frac{\nu \xi^{-2} (k_j-k_i)}{(k_j-k_i)^2+\xi^{-2}}DD' \nonumber\\
&\times
\Big\{
  GG'(\HomoPot_n+\HomoPot_n')
+ HG'(\HeteroPot_n+\tHeteroPot_n')\nonumber\\&
+ GH (\tHeteroPot_n+\HeteroPot_n')
+ HH (\tHomoPot_n+\tHomoPot_n')
\Big\}\\
\elabel{HeteroPot_recurrence}
\HeteroPot_{n+1}(k_j)=&(-k_j)\frac{1}{L}
    \sum_{
    \substack{i\in\Zset\\
    i\ne0}}
    \int\dintbar{\omega'} 
 \frac{\nu \xi^{-2} (k_j-k_i)}{(k_j-k_i)^2+\xi^{-2}}DD' \nonumber\\
&\nonumber\times
\Big\{
  HG'(\HomoPot_n+\HomoPot_n')
+ EG'(\HeteroPot_n+\tHeteroPot_n')\nonumber\\&
+ HH (\tHeteroPot_n+\HeteroPot_n')
+ EH (\tHomoPot_n+\tHomoPot_n')
\Big\}\\
\elabel{tHeteroPot_recurrence}
\tHeteroPot_{n+1}(k_j)=&(-k_j)\frac{1}{L}
    \sum_{
    \substack{i\in\Zset\\
    i\ne0}}
    \int\dintbar{\omega'} 
 \frac{\nu \xi^{-2} (k_j-k_i)}{(k_j-k_i)^2+\xi^{-2}}DD'\nonumber\\
&\nonumber\times
\Big\{
  GH (\HomoPot_n+\HomoPot_n')
+ HH (\HeteroPot_n+\tHeteroPot_n')\nonumber\\&
+ GE' (\tHeteroPot_n+\HeteroPot_n')
+ HE' (\tHomoPot_n+\tHomoPot_n')
\Big\}\\
\elabel{tHomoPot_recurrence}
\tHomoPot_{n+1}(k_j)=&(-k_j)\frac{1}{L}
    \sum_{
    \substack{i\in\Zset\\
    i\ne0}}
    \int\dintbar{\omega'} 
 \frac{\nu \xi^{-2} (k_j-k_i)}{(k_j-k_i)^2+\xi^{-2}}DD'\nonumber\\
&\nonumber\times
\Big\{
  HH (\HomoPot_n+\HomoPot_n')
+ EH (\HeteroPot_n+\tHeteroPot_n')\nonumber\\&
+ HE' (\tHeteroPot_n+\HeteroPot_n')
+ EE' (\tHomoPot_n+\tHomoPot_n')
\Big\} \ .
\end{align}
\end{subequations}
The zero-mode $k_i=0$ needs to be removed from the summations in \Erefs{PotsRecursion}, 
so as to correct for an artefact that arises
without a mass $r$ in the propagator, normally used to
regularise the infrared.
With a positive mass, it is clear that terms in the sum of the kind $k_i^2/(\diff k_i^2+r)$ vanish at $i=0$, but this
is ambiguous if $r=0$ as in $D$, \Eref{prop_conventions_D}.
This caveat is further discussed in Refs.~\cite{ZhangETAL:2024,ZhangGarcia-Millan:2023} 
in the context of the barometric formula.
The diagrammatic representation in \Eref{HomoPot_recurrence_diag}
corresponds to \Eref{HomoPot_recurrence}, and the diagrammatic representation of 
\Erefs{HeteroPot_recurrence},
\eref{tHeteroPot_recurrence}
and \eref{tHomoPot_recurrence}
are shown in
\Erefs{PotsRecursion_diag} in \aref{details_iteration}.

At the heart of the present work is a systematic way to calculate \Erefs{PotsRecursion}.
The details of the technical steps to do this are laid out in the following subsections.

\subsection{Reparametrisation}
The first step towards calculating the effective interaction vertices \Erefs{PotsAsSums} is identifying symmetries 
between them that derive from symmetries in the bare propagators in \Erefs{bare_props}
and the interaction vertices in \Erefs{vertices}. 
In \aref{symmetriesPhiPsi}, we demonstrate that the effective vertices satisfy
    \begin{subequations}
    \elabel{Pot_n_induction_condition}
        \begin{align}
    \elabel{HomoPot_n_induction_condition}
            \tHomoPot_n(k_j) =& \HomoPot_n(k_j) \ , \\
    \elabel{HeteroPot_n_induction_condition}
            \tHeteroPot_n(k_j) =& \HeteroPot_n(-k_j) \ ,
        \end{align}
    \end{subequations}
    for every $n$,
    reducing the total number of effective vertices from four in \Erefs{PotsAsSums} to two.
    In \aref{symmetriesPhiPsi}, we also show that both $\HomoPot(x)$ and $\HeteroPot(x)$ are real, and that
    $\HomoPot(x)$ is even in $x$.

The integral over $\omega$ in \Erefs{PotsRecursion}, \sref{omega_integrals}, will generally generate powers
of $k_j$ in the denominator which are important to understand as $|k_1|=2\pi/L$, the smallest $|k_j|$ in the
summations of \Erefs{PotsRecursion}, is arbitrarily small in large $L$.
It turns out that these negative powers of $k_j$ cancel neatly with the leading orders of
$\HomoPot_n$ and $\HeteroPot_n$, which we reparametrise as 
\begin{subequations}
\elabel{def_PQR_n}
\begin{align}
    \elabel{def_P_n}
    \HomoPot_n(k_j) =& k_j^2 P_n(k_j)\\
    \elabel{def_QR_n}
    \HeteroPot_n(k_j) =& k_j^2 Q_n(k_j) + \imag k_j R_n(k_j) \ ,
\end{align}
\end{subequations}
with the perspective that the real functions $P(k_j)$, $Q(k_j)$ and $R(k_j)$ are regular in small $k_j$, to be
confirmed below, as well as even.
The latter readily follows from physical reasoning: (Effective) potentials in real-space must be real, so that
$\HomoPot^*(k_j)=  \HomoPot(-k_j)$
and $\HeteroPot^*(k_j)=  \HeteroPot(-k_j)$.
The equal-species potential $\HomoPot(x)$ must be even (and real), so that its Fourier coefficients $\HomoPot(k_j)$
are also real and even, while $\HeteroPot(k_j)$ is expected to have an imaginary and odd contribution
for any finite self-propulsion. 
It follows that $P$ is real and even, that $Q$ and $R$ are even, given they are real, and further that the reparametrisation
\Erefs{def_PQR_n} does not pose a constraint on $\HomoPot_n$ and $\HeteroPot_n$. Instead of relying on the physical
properties of $\HomoPot$ and $\HeteroPot$, in the following we rederive these properties under the assumption of real
$P_n$, $Q_n$, $R_n$ on the basis of symmetries of $\HomoPot_n$ and $\HeteroPot_n$, 
\Erefs{tHomoPot_is_HomoPot_summary}, \eref{HomoPot_real} and \eref{HeteroPot_complex_is_odd},
which we derive by induction in \aref{symmetriesPhiPsi}.
Once $P_n$, $Q_n$, $R_n$ are established as even, real and well behaved in small $k$, they can themselves be
parametrised more efficiently, \Eref{PQR_reparam}.

Since $\HomoPot_n$ is even and real, \Erefs{tHomoPot_is_HomoPot_summary} and \eref{HomoPot_real}, 
it follows immediately that $P_n$ is even and real.
From  \Eref{HeteroPot_complex_is_odd_a}, we have that
\begin{equation}
    \HeteroPot_n(k_j)+\HeteroPot^*_n(k_j) = 
    \HeteroPot_n(k_j)+\HeteroPot_n(-k_j) \ ,
\end{equation}
so that the real part of $\HeteroPot_n$, namely $\realpart(\HeteroPot_n)=(\HeteroPot_n+\HeteroPot_n^*)/2$, is even in $k_j$ and, therefore, $Q_n$ is even and real.
Similarly, we have
\begin{equation}
    \HeteroPot_n(k_j)-\HeteroPot^*_n(k_j) = 
    \HeteroPot_n(k_j)-\HeteroPot_n(-k_j) \ ,
\end{equation}
so that the imaginary part of $\HeteroPot_n$, namely $\imaginarypart(\HeteroPot_n)=(\HeteroPot_n-\HeteroPot_n^*)/2$, 
is odd in $k_j$,
and thus $R_n$ is even and real.
Using \eref{HeteroPot_complex}, we can write $Q_n$ and $R_n$ as
\begin{subequations}
\begin{align}
2Q_{n}k_j^2	
=&
\HeteroPot_{n}+\tHeteroPot_{n} \ ,\\
2R_{n}\imag k_j 
=&
\HeteroPot_{n}-\tHeteroPot_{n} \ .
\end{align}
\end{subequations}
Using the following relations derived from \Erefs{Pot_n_induction_condition}, \eref{def_PQR_n},
\eref{tHomoPot_is_HomoPot} 
and \eref{tHeteroPot_is_HeteroPot}
\begin{subequations}
\elabel{def_PQR}
\begin{align}
\HomoPot_n+\HomoPot_n'=&\tHomoPot_n+\tHomoPot_n'=2\HomoPot_n=2k_j^2P_n\\
\HeteroPot_n+\tHeteroPot_n'=&2\HeteroPot_n=2(k_j^2Q_n+\imag k_j R_n)\\
\tHeteroPot_n+\HeteroPot_n'=&2\tHeteroPot_n=2(k_j^2Q_n-\imag k_j R_n)
\end{align}
\end{subequations}
in \Eref{PotsRecursion}, we obtain the recurrence relation for $P$, $Q$ and $R$,
\begin{subequations}
\elabel{PQR_full_integral}
\begin{align}
P_{n+1}(k_j)k_j^2 =&(-k_j)\frac{1}{L}
    \sum_{
    \substack{i\in\Zset\\
    i\ne0}}
    \int\dintbar{\omega'} 
\frac{{\nu \xi^{-2} (k_j-k_i)}}{(k_j-k_i)^2+\xi^{-2}}DD'
\nonumber\\
& 
\begin{aligned}
\times
\Big\{
  &\phantom{+\ }
   2 GG' k_i^2 P_n(k_i) \\
&+  2 HG'(k_i^2 Q_n(k_i) + \imag k_i R_n(k_i)) \\
&+  2 GH (k_i^2 Q_n(k_i) - \imag k_i R_n(k_i)) \\
&+  2 HH k_i^2 P_n(k_i)
\Big\}
\end{aligned}
\elabel{P_nP1_full_integral}
\\
2Q_{n+1}(k_j)k_j^2\elabel{Q_nP1_full_integral}
= &(-k_j)\frac{1}{L}
    \sum_{
    \substack{i\in\Zset\\
    i\ne0}}
    \int\dintbar{\omega'} 
\frac{{\nu \xi^{-2} (k_j-k_i)}}{(k_j-k_i)^2+\xi^{-2}}DD' \nonumber\\
&\,
\begin{aligned}
\times \Big\{
  &\phantom{+\ }
  2 (HG'+GH) k_i^2 P_n(k_i)\\
&+ 2 (EG'+HH)(k_i^2 Q_n(k_i) + \imag k_i R_n(k_i))\\
&+ 2 (HH+GE') (k_i^2 Q_n(k_i) - \imag k_i R_n(k_i))\\
&+ 2 (EH+HE') k_i^2 P_n(k_i)
\Big\}
\end{aligned}
\\
2R_{n+1}(k_j)\imag k_j	\elabel{R_nP1_full_integral}
=&(-k_j)\frac{1}{L}
    \sum_{
    \substack{i\in\Zset\\
    i\ne0}}
    \int\dintbar{\omega'} 
\frac{{\nu \xi^{-2} (k_j-k_i)}}{(k_j-k_i)^2+\xi^{-2}}DD' \nonumber\\
&\,
\begin{aligned}
\times\Big\{
  &\phantom{+\ }
  2 (HG'-GH) k_i^2 P_n(k_i)\\
&+ 2 (EG'-HH)(k_i^2 Q_n(k_i) + \imag k_i R_n(k_i))\\
&+ 2 (HH-GE') (k_i^2 Q_n(k_i) - \imag k_i R_n(k_i))\\
&+ 2 (EH-HE') k_i^2 P_n(k_i)
\Big\} \ .
\end{aligned}
\end{align}
\end{subequations}
The remaining challenge is to show that $P$, $Q$ and $R$ are regular in small $k_j$.
This amounts to showing that sum and integral in \Erefs{P_nP1_full_integral} and \eref{Q_nP1_full_integral}
vanish at least linearly in $k_j$ and that they are regular in small $k_j$ for \Eref{R_nP1_full_integral}.
We will show this by explicitly evaluating the right-hand sides of \Erefs{PQR_full_integral}.

\subsection{Loop integral over $\omega$}
\seclabel{omega_integrals}
The iterative step in \Eref{PQR_full_integral} involves an integral over $\omega$
that can be written as linear combinations of
\begin{subequations}
\elabel{ints_omega_def}
\begin{align}
    \Omega_0 =  & \int \dintbar{\omega} DD' \ ,\\
        \elabel{omega1int_def}
    \Omega_1= & \int \dintbar{\omega} DD' \omega \ ,\\
    \Omega_2= &
    \int \dintbar{\omega} DD' \omega^2 \ .
\end{align}
\end{subequations}
As $DD'$ is even in $\omega$, \Erefs{prop_conventions},
then the integrand of \eref{omega1int_def} is odd in $\omega$ and 
it follows immediately that $\Omega_1=0$.
The four simple poles of $DD'$, namely
$\pm\imag a_1$ and $\pm\imag a_2$ with
\begin{equation}\elabel{def_a12}
    a_1=\sqrt{\frac{\gamma}{\diff}}
    \quad\text{ and }\quad
    a_2=\sqrt{2\frac{\gamma}{\diff}+\frac{\drift^2}{\diff^2}} 
\end{equation}
determine the integrals \Erefs{ints_omega_def} producing
\begin{subequations}
\elabel{ints_omega}
\begin{align}
    \elabel{omega0int}
    \Omega_0 =  & 
    \frac{1}{4\diff^3 k_j^2} \frac{1}{a_2^2 - a_1^2} \left(
    \frac{1}{k_j^2+a_1^2} - \frac{1}{k_j^2+a_2^2} 
    \right) \ ,\\
        \elabel{omega1int}
    \Omega_2= &
    \frac{1}{4\diff} \frac{1}{k_j^2+a_1^2} \ .
\end{align}
\end{subequations}
The value of $\Omega_0$ is well defined because
the term $a_2^2 - a_1^2={\gamma}/{\diff}+{\drift^2}/{\diff^2}>0$ for any $\gamma/\diff>0$ or 
$\drift,\diff\neq0$.
Substituting $G,E,H,D$ \eref{prop_conventions} and 
the integrals $\Omega_0$, $\Omega_1$, $\Omega_2$ into
 \Eref{PQR_full_integral}
gives
\newcommand{\pp}{\phantom{+}\,}
\begin{align}\elabel{P_nP1_k_integral}
P_{n+1}(k_j)k_j^2=&(-k_j)\frac{1}{L}
    \sum_{
    \substack{i\in\Zset\\
    i\ne0}}
\frac{{\nu \xi^{-2} (k_j-k_i)}}{(k_j-k_i)^2+\xi^{-2}} \nonumber
\\
&\ \times
\Big\{
\begin{aligned}[t]
 &\left[\varphi_{P0} + \frac{\varphi_{P1}}{k_i^2+a_1^2} + \frac{\varphi_{P2}}{k_i^2+a_2^2} \right] P_n(k_i)\\
+&\left[\varphi_{Q0} + \frac{\varphi_{Q1}}{k_i^2+a_1^2} + \frac{\varphi_{Q2}}{k_i^2+a_2^2} \right] Q_n(k_i)\\
+&\left[\varphi_{R0} + \frac{\varphi_{R1}}{k_i^2+a_1^2} + \frac{\varphi_{R2}}{k_i^2+a_2^2} \right] R_n(k_i)
\Big\} \ ,
\end{aligned}
\end{align}
with coefficients derived in \Eref{def_f},
\begin{subequations}
\elabel{P_coeffs}
\begin{align}
 \varphi_{P0}=&\frac{1}{\diff} \ ,
&\varphi_{Q0}=&0 \ ,
&\varphi_{R0}=&0 \ ,
\\
 \varphi_{P1}=&- \frac{\gamma\drift^2}{\diff^2} \frac{1}{\gamma\diff+\drift^2} \ ,
&\varphi_{Q1}=&0 \ ,
&\varphi_{R1}=&-\frac{\drift\gamma}{\diff}\frac{1}{\gamma\diff+\drift^2} \ ,
\\
 \varphi_{P2}=&- \frac{1}{\diff} \frac{\gamma^2}{\gamma\diff+\drift^2} \ ,
&\varphi_{Q2}=&\frac{\gamma}{\diff^2}  \ ,
&\varphi_{R2}=&\frac{\drift\gamma}{\diff}\frac{1}{\gamma\diff+\drift^2} 
\ .
\end{align}
\end{subequations}

Similarly,
\begin{align}\elabel{Q_nP1_k_integral}
Q_{n+1}(k_j)k_j^2=&(-k_j)\frac{1}{L}
    \sum_{
    \substack{i\in\Zset\\
    i\ne0}}
\frac{{\nu \xi^{-2} (k_j-k_i)}}{(k_j-k_i)^2+\xi^{-2}} \nonumber
\\
&\ \times
\Big\{
\begin{aligned}[t]
 &\left[\gamma_{P0} + \frac{\gamma_{P1}}{k_i^2+a_1^2} + \frac{\gamma_{P2}}{k_i^2+a_2^2} \right] P_n(k_i)\\
+&\left[\gamma_{Q0} + \frac{\gamma_{Q1}}{k_i^2+a_1^2} + \frac{\gamma_{Q2}}{k_i^2+a_2^2} \right] Q_n(k_i)\\
+&\left[\gamma_{R0} + \frac{\gamma_{R1}}{k_i^2+a_1^2} + \frac{\gamma_{R2}}{k_i^2+a_2^2} \right] R_n(k_i)
\Big\}
\end{aligned}
\end{align}
with coefficients derived in \Eref{def_g}
\begin{subequations}
\elabel{Q_coeffs}
\begin{align}
 \gamma_{P0}=&0 \ ,
&\gamma_{Q0}=&\frac{1}{\diff} \ ,
&\gamma_{R0}=&0 \ ,
\\
 \gamma_{P1}=&0 \ ,
&\gamma_{Q1}=&0 \ ,
&\gamma_{R1}=&0 \ ,
\\
 \gamma_{P2}=&\frac{\gamma}{\diff^2} \ ,
&\gamma_{Q2}=&-\frac{\gamma\diff+\drift^2}{\diff^3} \ ,
&\gamma_{R2}=&\frac{-\drift}{\diff^2} 
\ ,
\end{align}
\end{subequations}
and
\begin{align}\elabel{R_nP1_k_integral}
R_{n+1}(k_j)\imag k_j=&(-k_j)\frac{1}{L}
    \sum_{
    \substack{i\in\Zset\\
    i\ne0}}
\frac{{\nu \xi^{-2} (k_j-k_i)}}{(k_j-k_i)^2+\xi^{-2}}
\nonumber\\
&\ \times
\imag k_i
\Big\{
\begin{aligned}[t]
 &\left[\eta_{P0} + \frac{\eta_{P1}}{k_i^2+a_1^2} + \frac{\eta_{P2}}{k_i^2+a_2^2} \right] P_n(k_i)\\
+&\left[\eta_{Q0} + \frac{\eta_{Q1}}{k_i^2+a_1^2} + \frac{\eta_{Q2}}{k_i^2+a_2^2} \right] Q_n(k_i)\\
+&\left[\eta_{R0} + \frac{\eta_{R1}}{k_i^2+a_1^2} + \frac{\eta_{R2}}{k_i^2+a_2^2} \right] R_n(k_i)
\Big\} \ ,
\end{aligned}
\end{align}
with coefficients derived in \Eref{def_h}
\begin{subequations}
\elabel{R_coeffs}
\begin{align}
& \eta_{P0}=0 \ ,\
\eta_{Q0}=0 \ ,\
\eta_{R0}=0 \ ,
\\
& \eta_{P1}=\frac{\gamma\drift}{\diff(\gamma\diff+\drift^2)} \ ,\
\eta_{Q1}=0 \ ,\
\eta_{R1}=\frac{\gamma}{\gamma \diff + \drift^2} \ ,
\\
& \eta_{P2}=-\frac{\gamma\drift}{\diff(\gamma\diff+\drift^2)} \ ,\
\eta_{Q2}=\frac{\drift}{\diff^2} \ ,\
\eta_{R2}=\frac{\drift^2}{\diff(\gamma \diff + \drift^2)} 
\ .
\end{align}
\end{subequations}

To proceed further, we introduce a reparameterisation of the even functions $P$, $Q$ and $R$,
\begin{subequations}
\elabel{PQR_reparam}
\begin{align}
\elabel{P_reparam}
    P_n(k_j) =& 
    {(\nu\xi^{-2})^n}
    \sum_{m=1}^{M_n} \pi_{n,m} \frac{1}{k_j^2+p^2_{n,m}} \ ,\\
\elabel{Q_reparam}
    Q_n(k_j) =& 
    {(\nu\xi^{-2})^n}
    \sum_{m=1}^{M_n} \zeta_{n,m} \frac{1}{k_j^2+p^2_{n,m}}\ , \\
\elabel{R_reparam}
    R_n(k_j) =& 
    {(\nu\xi^{-2})^n}
    \sum_{m=1}^{M_n} \rho_{n,m} \frac{1}{k_j^2+p^2_{n,m}} \ ,
\end{align}
\end{subequations}
with simple poles at $\pm\imag p_{n,m}$.
The number of those poles at any order $n$ of the iterative scheme \Erefs{P_nP1_k_integral}, 
\eref{Q_nP1_k_integral} and \eref{R_nP1_k_integral}
as well as their values are determined below.
The parametrisation in \Eref{PQR_reparam} implicitly assumes that $P$, $Q$ and $R$ 
have simple poles only. As it is shown below, this is not the case because double and higher-order
poles are generated under the iteration. However, we show below that the amplitude multiplying them is
exponentially, $\propto\exp{-L/\xi}$, small.
From \Erefs{pots1} and \eref{def_PQR_n}, we have
\begin{subequations}
\elabel{def_PQR1}
\begin{align}
    P_1(k_j) =& -\frac{{\nu \xi^{-2}}}{2L^2} \frac{1}{k_j^2+\xi^{-2}}\\
    Q_1(k_j) =& -\frac{{\nu \xi^{-2}}}{2L^2} \frac{1}{k_j^2+\xi^{-2}}\\
    R_1(k_j) =&  0 \ ,
\end{align}
\end{subequations}
so that 
\begin{equation}
\elabel{pole_11}
    p_{1,1}=\xi^{-1} \ ,
\end{equation} 
and
\begin{align}
\elabel{pi_zeta_rho_n1}
    \pi_{1,1}  = 
    -\frac{1}{2L^2} , \ &&
    \zeta_{1,1}  = 
    -\frac{1}{2L^2} , \ &&
    \rho_{1,1} = 0 \ .
\end{align}
In the following, we adopt the language of ``pole $p_{n,m}$" and ``contributions to pole $p_{n,m}$"
when we mean poles at $\pm\imag p_{n,m}$ and contributions to $\hat{A}$ in a term of the form
$\hat{A}/(k_j^2+p^2_{n,m})$ respectively, which has residues $\pm\hat{A}/(2\imag p_{n,m})$.
The ``amplitude" $\hat{A}$ is $\pi_{n,m}$, $\zeta_{n,m}$ and $\rho_{n,m}$ respectively in \Erefs{PQR_reparam}.
The iterative scheme \Erefs{P_nP1_k_integral}, \eref{Q_nP1_k_integral} and \eref{R_nP1_k_integral}
with coefficients in \Erefs{P_coeffs}, \eref{Q_coeffs} and \eref{R_coeffs} can be found explicitly, in dimensionless
form, in \aref{explicit_iteration_PQR}.

\subsection{Loop sum over $k_j$}
Substituting $P_n$, $Q_n$, $R_n$ in \Erefs{PQR_reparam} into the iterative scheme in
\Erefs{P_nP1_k_integral}, \eref{Q_nP1_k_integral} and \eref{R_nP1_k_integral}
gives the loop sums in $k_j$ that are needed to derive the $(n+1)$-th order.
These sums are generally convolutions, with one term of the form $((k_j-k_i)^2+\xi^{-2})^{-1}$ multiplying terms of the form $(k_i^2+\alpha)^{-1}$, either in the square brackets of \Erefs{P_nP1_k_integral}, \eref{Q_nP1_k_integral} and \eref{R_nP1_k_integral} or in the $P_n$, $Q_n$, $R_n$, \Erefs{PQR_reparam}, multiplying them. Specifically, the coefficients with index $0$, such as $\varphi_{P0}$, $\varphi_{Q0}$ and $\varphi_{R0}$ in \Eref{P_nP1_k_integral}, and similarly $\gamma_{P0}$ \etc in \Eref{Q_nP1_k_integral}, multiply only the fractions $(k_i^2+\alpha)^{-1}$ in \Eref{PQR_reparam}, whereas the coefficients $\varphi_{P1}$, $\varphi_{Q1}$, $\varphi_{R1}$ and $ \varphi_{P2}$ \etc multiply products of the form $(k_i^2+\alpha)^{-1}(k_i^2+\beta)^{-1}$, with one term explicitly in the square brackets and one in the $P_n$, $Q_n$, $R_n$, \Erefs{PQR_reparam}. To bring these products $(k_i^2+\alpha)^{-1}(k_i^2+\beta)^{-1}$ into the form of sums of $(k_i^2+\alpha)^{-1}$ and $(k_i^2+\beta)^{-1}$ with suitable amplitudes, we write them for $\alpha\ne\beta$ as partial fractions,
\begin{equation}
\elabel{partial_fractions}
    \frac{1}{k_j^2+\alpha^2} \frac{1}{k_j^2+\beta^2} = \frac{1}{\alpha^2-\beta^2}
    \left(
    \frac{1}{k_j^2+\beta^2} - \frac{1}{k_j^2+\alpha^2} 
    \right) \ .
\end{equation}
This reduces all sums to
the following loop sums over $k_i$, which are known as Matsubara sums
\cite{Matsubara:1955,Espinosa:2010},
and are
derived in \aref{app_sums} by applying the summation theorem 
\cite{SpiegelETAL:2009,MarsdenHoffman:1973},
\begin{subequations}
\elabel{ksum}
\begin{align}
    \frac{1}{L}
    \sum_{
    \substack{i\in\Zset\\
    i\ne0}}
    &
    \frac{k_j-k_i}{(k_j-k_i)^2+\xi^{-2}} \frac{1}{k_i^2+\alpha^2}
    \nonumber\\
     = &\frac{k_j}{2\alpha}
     \left(
 \frac{\Aa(\alpha)}{k_j^2+(\alpha+\xi^{-1})^2}
+
 \frac{\Ba(\alpha)}{k_j^2+(\alpha-\xi^{-1})^2}
 \right.\nonumber\\ &\left.
-
 \frac{2}{L \alpha}\frac{1}{k_j^2+\xi^{-2}}
 \right)
    \elabel{k0sum}
\\
    \frac{1}{L}
    \sum_{
    \substack{i\in\Zset\\
    i\ne0}}
    &
    \frac{k_j-k_i}{(k_j-k_i)^2+\xi^{-2}} \frac{k_i}{k_i^2+\alpha^2} \nonumber\\
    \elabel{k1sum}
    & = -\frac{1}{2}
 {\left(
   \frac{\Ab(\alpha) (\alpha+\xi^{-1})}{k_j^2+(\alpha+\xi^{-1})^2} 
+ \frac{\Bb(\alpha) (\alpha-\xi^{-1})}{k_j^2+(\alpha-\xi^{-1})^2} 
 \right)}
    \ ,
\end{align}
\end{subequations}
with coefficients
\begin{subequations}
\elabel{def_AB}
\begin{align}
   \Ab(\alpha)=&\frac{1-\exp{-L(\alpha+\xi^{-1})}} {(1-\exp{-L\alpha})(1-\exp{-L/\xi})} \ , \elabel{ampA}\\
   \Bb(\alpha)=&\frac{\exp{-L\alpha}-\exp{-L/\xi}} {(1-\exp{-L\alpha})(1-\exp{-L/\xi})} \ . \elabel{ampB}
\end{align}
\end{subequations}
Of these, $A(\alpha)$ may be thought of as ``order $1$" and $B(\alpha)$ may be thought of as vanishing, because for $L\alpha$ large, as we will assume, and $L\xi^{-1}$ large anyway,
the amplitudes are $A(\alpha) = 1+\OC(\exp{-L\alpha})$ and $B(\alpha) = \OC(\exp{-L\alpha})$, \Erefs{AB_L_limit}.
For $L\gg\alpha$, the exponentials essentially vanish.
Although we keep track of terms proportional to both $A(\alpha)$ and $B(\alpha)$
we will draw on the fact that $B(\alpha)$ is small when considering some of the poles, 
in particular repeated ones. These are not covered by \Eref{partial_fractions},
as it assumes $\alpha\neq\beta$, and are difficult to consider in full generality.
The loop sum in \eref{k0sum} features in $P$ and $Q$, \Erefs{P_nP1_k_integral} and \eref{Q_nP1_k_integral},
whereas the loop sum in \eref{k1sum} 
features in $R$, \Eref{R_nP1_k_integral}, as the curly bracket is multiplied by $k_i$.

\subsection{Iterative generation of poles}

\begin{figure*}
\centering
\includegraphics[width=0.28\textwidth]{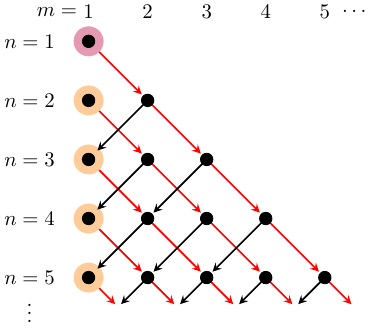}
\qquad
\qquad
\includegraphics[width=0.463\textwidth]{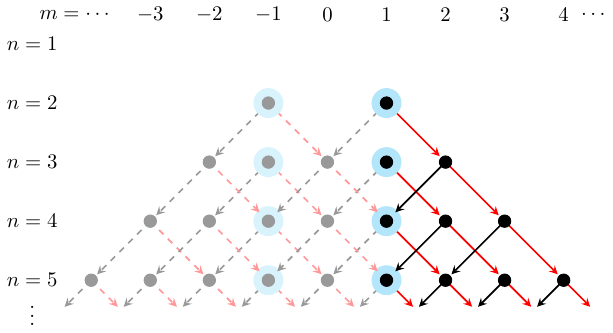}
\caption{\flabel{fig_poles}
Pole generation $p_{n,m}^{(i)}$ order by order (left) $\xi$-type $p_{n,m}^{(0)}=m\xi^{-1}$
and (right) $a_i$-type $p_{n,m}^{(i)}=a_i+m\xi^{-1}$, \Eref{pole_types}.
Arrows indicate up-promotion with amplitude $A$, \Eref{ampA} (red),
and down-promotion with amplitude $B$, \Eref{ampB} (black).
Base case $p_{1,1}^{(0)}=\xi^{-1}$, \Eref{pole_11}, is encircled in purple.
Poles of $\xi$-type encircled on the left
in orange are generated at every order through the final term in \Eref{k0sum} $\sim L^{-1}$,
which does not carry a factor of $A$ or $B$.
The poles labelled blue on the right are populated at every order from $a_i$-poles in the square brackets in
\Erefs{P_nP1_k_integral}, \eref{Q_nP1_k_integral} and \eref{R_nP1_k_integral},
which via \Eref{partial_fractions} are shifted by \Eref{ksum} to $a_i\pm\xi^{-1}$, so that
$p_{n,-1}^{(i)}=a_i-\xi^{-1}$ carries a factor of $B(a_i)$ and 
$p_{n,1}^{(i)}=a_i+\xi^{-1}$ a factor of $A(a_i)$.
The orange and blue labelled poles are the only ones freshly populated in the scheme, the amplitude of all others
``inherit" the amplitudes of those in previous generations.
As a result, all $a_i$-type poles for $m\leq0$ (faded) carry a factor of $B$
and are thus omitted in the present scheme.
Double poles generated from $p_{n,0}^{(i)}=a_i$ are also omitted.
There is no shift below $p_{n,1}^{(0)}=\xi^{-1}$ on the left because of \Eref{B_xi_zero}.
}
\end{figure*}

The loop sums in \Eref{ksum} reveal how the poles in $k_j$ arise iteratively
as we proceed from $n$ loops to $n+1$ loops.
We find that every pole $\alpha$ that exists to $n$ loops is shifted ``to the right" by $\xi^{-1}$
(up-promotion)
as well as ``to the left" by $\xi^{-1}$ (down-promotion), generating in the next order the new poles 
$\alpha+\xi^{-1}$ and $\alpha-\xi^{-1}$ respectively.
The amplitude with which they contribute to the next order is proportional to the functions $A$ and $B$ respectively, showing that the ``down-promotion" of poles has exponentially vanishing amplitude for large system sizes 
$L\gg\xi$, $a_i^{-1}$, as discussed in detail below.
The algebraic term in $L$ in
\Eref{k0sum} further shows an additional contribution to the pole $\xi^{-1}$ at every order $n$.
In this iterative generation of poles, \Eref{k0sum} seems to produce a pole at $k_j=0$ through $B$.
However,
the special case of $\alpha=\xi^{-1}$, 
which produces the term
\begin{equation}
\elabel{B_xi_zero}
\frac{k_j}{2\xi^{-1}} \frac{B(\xi^{-1})}{k_j^2} = 0
\end{equation}
is resolved by a zero amplitude
$B(\xi^{-1})=0$ as given by \Eref{ampB}.

At first order, $n=1$, there is actually only one pole, namely $\xi^{-1}$ in \eref{pole_11}.
However, for better book keeping it pays to consider two more poles,
$a_1$ and $a_2$, \Erefs{def_a12},
even when they feature with vanishing amplitude at $n=1$.
Via \Erefs{partial_fractions} and \eref{ksum},
the poles enter at order $n=2$ as $a_{1,2}\pm\xi^{-1}$.
We refer to these poles as different types, ``$\xi$-type", ``$a_1$-type" and ``$a_2$-type",
and use a superscript for quantities related to each pole type.
The set of poles at each order $n$, generated through \eref{ksum} is covered by
\begin{subequations}
\elabel{pole_types}
\begin{align}
\elabel{pole_type_xi}
    p^{(0)}_{n,\mtilde}=&m\xi^{-1} && \text{for } m\in\{1,\ldots,n\} \ ,\\
\elabel{pole_type_a1}
    p^{(1)}_{n,m}=&a_1+m\xi^{-1} && \text{for } m\in\{-(n-1),\ldots,n-1\} \ ,\\
\elabel{pole_type_a2}
    p^{(2)}_{n,m}=&a_2+m\xi^{-1} && \text{for } m\in\{-(n-1),\ldots,n-1\} \ ,
\end{align}
\end{subequations}
where amplitudes for $m\leq0$ of $a_i$-type poles in \eref{pole_type_a1} and \eref{pole_type_a2}
are treated as $0$, as discussed below.
Counting all poles, there are $M_n=5n-2$ poles at order $n$, of which $2n$ are treated as having vanishing amplitude.

\fref{fig_poles} illustrates the mechanisms by which poles are generated iteratively.
The shift by $\pm\xi^{-1}$ produces the same poles by different paths, resulting in different contributions
to its amplitude. For example, the pole $a_1+\xi^{-1}$ is generated 
by shifting $a_1$ by $+\xi^{-1}$ and by shifting $a_1+2\xi^{-1}$ by $-\xi^{-1}$.

It is cumbersome, however, to determine the different contributions to poles in one
generation from the poles in the previous generation. For example, the pole $a_1+\xi^{-1}$ in generation $n+1$ 
may have contributions from $a_1$ in the square brackets of 
\Erefs{P_nP1_k_integral}, \eref{Q_nP1_k_integral} and \eref{R_nP1_k_integral}
via \Eref{partial_fractions}, but from $a_1+2\xi^{-1}$ only if this pole exists in the 
previous generation $n$, \ie provided $n-1\ge2$. Finding these conditions retrospectively
is an arduous task.
Rather than writing down an explicit sum over all contributions to a given new pole, 
it is therefore more efficient to assign a given contribution to a particular residue in the next generation.
Similarly, it is not necessary to keep track of $a_i$-type poles that are not populated at a given generation,
it suffices to assign them a vanishing amplitude.

Neglecting double and higher-order poles simplifies the iterative calculation.
Double poles are generated amongst
$a_i$-type poles with $m=0$ at orders $n\geq3$ through 
shifts of poles at $a_i\pm\xi^{-1}$. Those at $a_i-\xi^{-1}$ are created with an amplitude 
$B(a_i)=\OC(\exp{-La_i})$ and are thus exponentially suppressed before shifted to $a_i$ with a factor
$A(a_i-\xi^{-1}) = 1+\OC(\exp{-L(a_i-\xi^{-1})})$.
Those at $a_i+\xi^{-1}$ are created with $A(a_i) = 1+\OC(\exp{-La_i})$ and subsequently multiplied by
$B(a_i+\xi^{-1})=\OC(\exp{-L(a_i+\xi^{-1})})$.
Amplitudes of repeated poles are therefore exponentially suppressed provided $L\xi^{-1}\gg1$ as well as
$La_1\gg1$ and $La_2\gg1$, \Eref{ampB}, or equally $L^2\gamma/\diff\gg1$, \Eref{def_a12}.
Making this simplifying assumption is a matter of pragmatism.
Sums such as \Eref{ksum} can be determined for a double pole $\alpha$, for example by differentiation
with respect to $\alpha$. Their results generate higher-order repeated poles, which are, in turn,
propagated through up/down-promotion.
To circumvent the generation of a cascade of higher-order poles, we ignore poles at
$\pm\imag p^{(i)}_{n,0}$ for $n\geq3$.
As all $a_i$-type poles for $m\leq0$ have at least one exponentially small factor of $B$ we ignore all of those
in our scheme, which thus requires $L/\xi\gg1$, as well as $L^2\gamma/\diff\gg1$, \Eref{def_a12}.

In summary, our iterative scheme follows from substituting the loop sums in \Erefs{ksum} into
\Erefs{P_nP1_k_integral}, \eref{Q_nP1_k_integral} and \eref{R_nP1_k_integral}
and
collecting all contributions to order $n+1$ from each pole at order $n$.
Instead of writing the explicit form of the effective interaction vertices 
$\HomoPot$, $\tHomoPot$, $\HeteroPot$ and $\tHeteroPot$ \eref{def_PQR_n},
or equivalently $P$, $Q$ and $R$ \eref{PQR_reparam},
we identify the contribution from each term in $P$, $Q$ and $R$ at oder $n$ to the coefficients
$\pi_{n+1,m}$,  $\zeta_{n+1,m}$ and $\rho_{n+1,m}$ at order $n+1$ with the initial values
in \Eref{pi_zeta_rho_n1} for $\xi$-type poles,
and $\pi^{(i)}_{n,m} = \zeta^{(i)}_{n,m} = \rho^{(i)}_{n,m} = 0$ for $a_i$-type poles.
The remaining algebra is detailed in Apps.~\ref{iteration_spelled_out} and 
\ref{iteration_potentials_pi_zeta_rho}.

\subsection{Sums vs integrals}
\seclabel{sum_vs_int}
In the large $L$ limit, the Riemann sums in \Eref{ksum} over discrete Fourier modes $k_i$ converge to 
integrals over continuous $k'$,
\begin{equation}
\lim_{L\to\infty}    \frac{1}{L} 
\sum_{i\in\Zset}
     f(k_i) =  \int\dintbar{k'} f(k')
\end{equation}
for arbitrary $f(k)$.
The loop integrals that approximate the sums in \Eref{ksum} are
\begin{subequations}
\elabel{kint_real_general}
\begin{align}
    \int \dintbar{k'} \frac{k-k'}{(k-k')^2+\xi^{-2}} \frac{1}{k'^2+\alpha^2} =& \frac{k}{2\alpha} \frac{1}{k^2+(\alpha+\xi^{-1})^2}\\
    \int \dintbar{k'} \frac{k-k'}{(k-k')^2+\xi^{-2}} \frac{k'}{k'^2+\alpha^2} =& 
    -\half\frac{\alpha+\xi^{-1}}{k^2+(\alpha+\xi^{-1})^2}\ ,
\end{align}
\end{subequations}
for real and positive $\alpha$, such as $\alpha = p^{(0)}_{1,1}=\xi^{-1}$ or $\alpha = p^{(i)}_{1,0}=a_i$.
With continuous $k$, the generation of poles has a much simpler structure
since the only mechanism at play is the ``up-promotion"
of any pole $\alpha$ by $\xi^{-1}$ to the right. 
As poles $a_i$ are populated at every order by \Eref{partial_fractions}, the set of poles reduces to
$2n-1$, namely
$p_{n}^{(0)}=n\xi^{-1}$ and $p_{n}^{(i)}=a_i+m\xi^{-1}$ for $m\in\{1,\ldots,n-1\}$.
In \fref{fig_poles}, all black arrows vanish in this limit, as well as $\xi$-type poles encircled in orange and
poles generated from them.

The coefficients 
$A(\alpha)$ 
and 
$B(\alpha)$ 
in \Eref{def_AB}
have limits
\begin{subequations}
\elabel{AB_L_limit}
\begin{align}
    \lim_{L\to\infty} \Ab(\alpha) = & 1 \ ,
\\
    \lim_{L\to\infty} \Bb(\alpha) = &0 \ ,
\end{align}
\end{subequations}
so that taking this limit on the right of \Eref{ksum} recovers
the loop integrals \Eref{kint_real_general} for fixed $k_j=k$.
While finite-size corrections contained in $A$ and $B$ in \Eref{def_AB} decay exponentially,
the third term in \Eref{k0sum} implements algebraic corrections in $L$ that originate from
the removal of the $0$-mode.
These algebraic corrections have a noticeable effect on the particle statistics, 
but are missed when simply replacing sums with integrals.

\section{Observables}
\seclabel{observables}

Having derived the effective interaction vertices $\HomoPot$ and $\HeteroPot$
\Erefs{PotsAsSums} and
\eref{def_PQR_n}
or, equivalently, $P$, $Q$ and $R$, calculating observables in the stationary state is a matter of
attaching the suitable outgoing legs to the effective vertices.
In the following, we use the dimensionless parameters
$\xibar = {\xi}/{L}$,
$\nubar = {\nu}/{(\diff\xi)}$,
$\Pe= {\drift^2}/{(\diff\gamma)}$, and
$\gammabar  = {\gamma \xi^2}/{\diff}$.

\subsection{Structure factor}
The static structure factor is defined as the Fourier transform of the pair correlation function $P(x_1,x_2)$ 
\cite{HansenMcDonald:2006}
\begin{equation}
\elabel{Sij_def}
S_{i,j}
= \int_{-L/2}^{L/2} \dint{x_1}\dint{x_2} \exp{-\imag (k_i x_1+k_j x_2)} P(x_1,x_2) \ .
\end{equation}
The stationary pair correlation function, or two-point particle number density, reads 
\begin{align}
\elabel{def_C}
        P(x_1,x_2)=  & \lim_{t_0\to-\infty}  \big\langle
(\phi(x_1,t) + \psi(x_1,t))
 \nonumber\\
&
(\phi(x_2,t) + \psi(x_2,t)) \phidagger(y_{1},t_{0})\phidagger(y_{2},t_{0})
\big\rangle \ .
\end{align}
It is the probability density to find one particle of any species at $x_1$ and another one at $x_2$,
so that its normalisation is $2$.
In steady state, it is a function only of the difference $x_1-x_2$, in fact it is even in the difference
and $L$-periodic in both arguments.
Attaching the bare propagators in \Eref{bare_props} as outgoing legs to the 
effective interaction vertices defined in \Eref{PotsAsSums}, 
the pair correlation function is
\begin{align}\elabel{C_k_algebra}
    S_{i,j} = &L \delta_{i+j,0} 
    \Big\{
    \frac{2}{L^2} L \delta_{i,0} + 
        \int \dintbar{\omega} DD'
\nonumber \\
    &\times
    \Big[
 \begin{aligned}[t]
     &(GG' + HG' + GH  + HH  )(\HomoPot   + \HomoPot') \\
    + &(HG' + EG' + HH + EH  )(\HeteroPot + \tHeteroPot')\\
    + &(GH  + HH + GE'+ HE' )(\HeteroPot' + \tHeteroPot)\\
    + &(HH + EH  + HE' + EE' )(\tHomoPot   + \tHomoPot')\Big]
    \Big\} \ .
 \end{aligned}
\end{align}
Defining $S_j=\sum_i S_{i,j} = S_{-j,j}$ and
substituting \Eref{def_PQR} into \eref{C_k_algebra} before taking the integral over $\omega$
using the functions $f_P$, $f_Q$, $f_R$,  $g_P$, $g_Q$, $g_R$ defined in \Erefs{def_f} and \eref{def_g}, 
we obtain
\begin{align}\elabel{C_k_algebra_final}
    S_{j}  
    =& 2\delta_{j,0} 
    + 2 L \sum_{n=1} 
    \Big[
    \begin{aligned}[t]
     &
       \big(f_P(k_j)+ f_Q(k_j)
       \big) P_n(k_j)\\
       +&\big( g_P(k_j)+ g_Q(k_j)         
       \big) Q_n(k_j)\\
       +&\big( f_R(k_j)+ g_R(k_j)       
              \big) R_n(k_j)
    \Big] \ .
    \end{aligned} 
\end{align}
Calculating the structure factor requires the evaluation of $P_n$, $Q_n$, $R_n$
iteratively up to a desired order $n$ in $\nu$, \Erefs{PQR_reparam}.
The structure factor $S_j$ is shown in \fref{structure_factor} as a function of $j$ for increasing activity.
In the passive case, $\Pe=0$, the structure factor $S_j$ is negative for all wavelengths $k_j$.
As the activity $\Pe$ increases, the lowest modes increase as well, eventually rendering $S_j$ positive.
The lowest mode, the compressibility factor $S_1$, is the fingerprint of effective attraction or effective repulsion
as introduced in \cite{letter}.
For $\xi\ll L$, so that $k_1\xi\ll1$, we obtain the following asymptotic behaviour in small
$\xibar=\xi/L$ of the compressibility $S_1$
by direct substitution of the coefficients in \Erefs{def_f} and \eref{def_g} in
\Eref{C_k_algebra_final},
\begin{align}
S_1 = &\frac{2L}{\diff(1+\frac{1}{2}\Pe)} \sum_{n\geq1}
\left(
P_n(0) + Q_n(0) - \sqrt{\frac{\Pe}{\gammabar}} \xi R_{n}(0) 
\right) 
\nonumber\\
&+\OC\left(\xibar^2\right)\ .
\elabel{C_k1_simple}
\end{align}
This simple expression, which features
the effective diffusivity $\mathsf{D}_{\text{eff}}=\diff(1+\Pe/2)$  in the denominator of $S_1$, 
captures the essence of how the effective attraction is encoded in the
perturbation expansion.
As discussed in \aref{explicit_iteration_PQR}, the sign of $P_n$ and $Q_n$ alternates between positive and 
negative for even and odd $n$, while $\xi R_n$ takes small numerical values compared to $P_n$ and $Q_n$.
Thus, the sign of each term in the expansion of $S_1$ essentially
follows the sign of $P_n$ and $Q_n$, while the amplitude of each term is reduced by $R_n$
depending on $\sqrt{\Pe/\gammabar}$.
This crucially affects odd terms in the expansion, which are the ones that carry the sign of
effective interactions since they are positive for bare attraction and negative for bare repulsion \cite{letter}.
Thus, odd terms in the expansion of $S_1$ are the only ones that can render $S_1$ negative due to 
bare repulsion.
The suppression of odd terms by $R_n$  in the expansion then leads to $S_1>0$
for large enough $\nubar$ and $\Pe>0$.
The compressibility $S_1$ is written explicitly up to second order in $\nubar$ in \aref{long_S1}.

\begin{figure}
\includegraphics[width=0.45\textwidth]{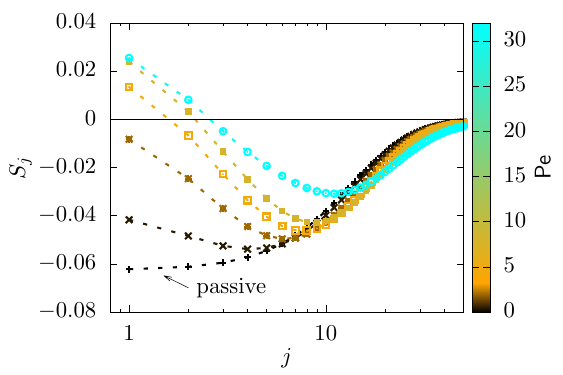}
\caption{\flabel{structure_factor} Structure factor $S_j$ as a function of Fourier mode $j$ for varying activity $\Pe=\drift^2/(\diff\gamma)$. 
Parameters: 
$\diff = 2$,
$L = 20$,
$\nubar = 5$,
$\xibar = 0.1$,
$\gammabar = 0.02$.}
\end{figure}

\subsection{Two-point correlation functions}
\seclabel{corr_funcs}

\begin{figure*}
\includegraphics[width=\textwidth]{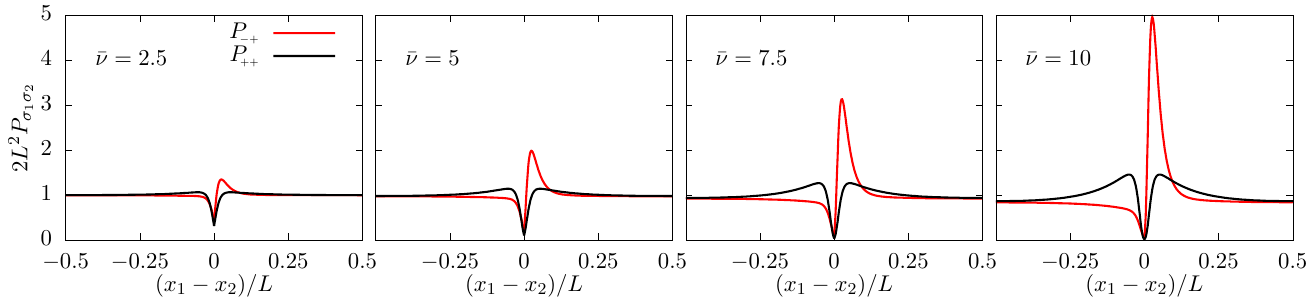}
\caption{\flabel{corr_funcs_nubar} 
Stationary two-point correlation functions $P_{++}$ and $P_{-+}$
\Erefs{stat_two_point_corrs} as a function of the inter-particle distance $x_1-x_2$,
for increasing interaction coupling $\nubar$.
Parameters correspond to Fig.~2(a) in \cite{letter}: 
$D = 0.5$, $L = 20$, $\Pe =20$, $\gammabar = 0.008$, $\xibar =  0.01$.
}
\end{figure*}

\begin{figure*}
\includegraphics[width=\textwidth]{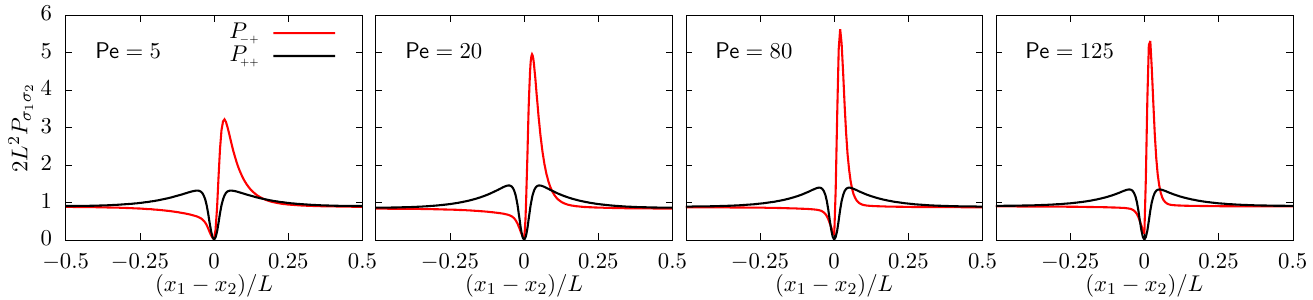}
\caption{\flabel{corr_funcs_Pe} 
Stationary two-point correlation functions $P_{++}$ and $P_{-+}$
\Erefs{stat_two_point_corrs} as a function of the inter-particle distance $x_1-x_2$,
for increasing activity $\Pe$.
Parameters correspond to Fig.~2(b) in \cite{letter}: 
$D = 0.5$, $L = 20$, $\nubar = 10$, $\gammabar=0.008$, $\xibar=0.01$.
}
\end{figure*}

The two-point correlation function $P(x_1,x_2)$, or two-point particle number density in \Eref{def_C},
is calculated by inverting
the Fourier transform \Eref{Sij_def} of the structure factor in \eref{C_k_algebra_final}.
Using \Eref{partial_fractions} and
\begin{equation}
\elabel{id_coshFourier}
 \frac{1}{L} \sum_{j=-\infty}^\infty \exp{\imag k_j x} \frac{1}{k_j^2+\alpha^2} 
    =  \ \frac{\cosh\left(\left(|x|-\frac{L}{2}\right)\alpha\right)}{2\alpha\sinh\left(\frac{L}{2}\alpha\right)} \ ,
\end{equation}
 the inverse Fourier transform of the structure factor
is a linear combination of \eref{id_coshFourier} where $\alpha$ takes the value of every pole
at each order in $\nu$. The inverse Fourier transform can thus be calculated exactly.
The two-point correlation function $P(x_1,x_2)$, which is shown in Fig.~2 of \cite{letter},
has normalisation such that
$\int_{-L/2}^{L/2} \dint{x_1} \dint{x_2}  P(x_1,x_2) = 2 $.

We also derive the steady-state joint probability densities 
\begin{subequations}
\elabel{stat_two_point_corrs}
\begin{align}
\Ppp(x_1,x_2)=  & \lim_{t_0\to-\infty}  \big\langle
\phi(x_1,t)
\phi(x_2,t) \phidagger(y_{1},t_{0})\phidagger(y_{2},t_{0})
\big\rangle \ , \\
\Pmp(x_1,x_2)=  & \lim_{t_0\to-\infty}  \big\langle
\psi(x_1,t)
\phi(x_2,t)\phidagger(y_{1},t_{0})\phidagger(y_{2},t_{0}) 
\big\rangle \ ,
\end{align}
\end{subequations}
which amounts to a calculation similar to the effective vertices 
$\HomoPot_{n+1}$ and $\HeteroPot_{n+1}$ in \Eref{PotsRecursion}.
Defining $\Pppj{i,j}=\delta_{i+j,0}$ and similarly $\Pmpj{i,j}$ as the Fourier transforms of \Erefs{stat_two_point_corrs}
in $x_1$, $x_2$ only, \Eref{Fourier_convention} without the integral over $t$, these observables can be expressed 
in terms of the Fourier-transformed fields, such as $\phi_{j}(\omega)$.
The steady state nature of \Erefs{stat_two_point_corrs} then results in an integral over $\omega$. Defining further
$\Pppj{j}=\sum_i \Pppj{i,j} = \Pppj{-j,j}$
and $\Pmpj{j}=\sum_i \Pmpj{i,j}= \Pmpj{-j,j}$,
we have
\begin{widetext}
\begin{subequations}
\elabel{Ppp_k_algebra}
\begin{align}
    \Pppj{j} = &
    \frac{1}{2}  \delta_{j,0} + 
       L \int \dintbar{\omega} DD' 
    \Big[
       GG' (\HomoPot   + \HomoPot') 
     + HG' (\HeteroPot + \tHeteroPot')
     +  GH  (\HeteroPot' + \tHeteroPot)
     + HH  (\tHomoPot   + \tHomoPot') 
    \Big]
    \\ 
        = &
    \frac{1}{2}  \delta_{j,0} + 
       L  \sum_{n=1} 
       \Big[
       \begin{aligned}[t]
  &  \left(\varphi_{P0} + \frac{\varphi_{P1}}{k_j^2+a_1^2} + \frac{\varphi_{P2}}{k_j^2+a_2^2}
   \right)  P_n(k_j)
   \\
  + & \left( \varphi_{Q0} + \frac{\varphi_{Q1}}{k_j^2+a_1^2} + \frac{\varphi_{Q2}}{k_j^2+a_2^2}
   \right)  Q_n(k_j)
   \\
  + & \left(\varphi_{R0} + \frac{\varphi_{R1}}{k_j^2+a_1^2} + \frac{\varphi_{R2}}{k_j^2+a_2^2}
      \right)  R_n(k_j)
   \Big]
    \end{aligned} 
\end{align}
\end{subequations}
and
\begin{subequations}
\elabel{Pmp_k_algebra}
\begin{align}
   \Pmpj{j} = &
    \frac{1}{2}  \delta_{j,0} + 
       L \int \dintbar{\omega} DD'
       \Big[
     HG' (\HomoPot   + \HomoPot') 
     +  EG' (\HeteroPot + \tHeteroPot')
     +  HH  (\HeteroPot' + \tHeteroPot)
     + EH  (\tHomoPot   + \tHomoPot')     \Big]
    \\ 
       = &
    \frac{1}{2}  \delta_{j,0} 
+ 
       L  \sum_{n=1} 
       \Big[
       \begin{aligned}[t]
&    \left(\gamma_{P0} + \imag k_j \eta_{P0} + \frac{\gamma_{P1}+ \imag k_j \eta_{P1}}{k_j^2+a_1^2} + \frac{\gamma_{P2}+ \imag k_j \eta_{P2}}{k_j^2+a_2^2}
   \right)  P_n(k_j)
   \\
  + & \left( \gamma_{Q0}+ \imag k_j \eta_{Q0} + \frac{\gamma_{Q1} + \imag k_j \eta_{Q1}}{k_j^2+a_1^2} + \frac{\gamma_{Q2} + \imag k_j \eta_{Q2}}{k_j^2+a_2^2}
   \right)  Q_n(k_j)
   \\
  + & \left(\gamma_{R0} + \imag k_j \eta_{R0} + \frac{\gamma_{R1} + \imag k_j \eta_{R1}}{k_j^2+a_1^2} + \frac{\gamma_{R2} + \imag k_j \eta_{R2}}{k_j^2+a_2^2}
      \right)  R_n(k_j)
   \Big] \ .
    \end{aligned} 
\end{align}
\end{subequations}
\end{widetext}
Using the identities of partial fractions in \Eref{partial_fractions}
and the periodic Yukawa potential in \Eref{id_coshFourier},
as well as  the derivative of the latter with respect to $x$,
\begin{align}
    \frac{1}{L} \sum_{j=-\infty}^\infty \exp{\imag k_j x} \frac{\imag k_j}{k_j^2+\alpha^2}  
    = \ {\sign(x)} \frac{\sinh\left(\left(|x|-\frac{L}{2}\right)\alpha\right)}{2\sinh\left(\frac{L}{2}\alpha\right)} \ ,
\end{align}
the correlation functions $\Ppp(x_1,x_2)$ and  $\Pmp(x_1,x_2)$ are calculated analytically in real space.
\fref{corr_funcs_nubar} shows the two-point correlation functions $\Ppp$ and $\Pmp$ for increasing interaction
coupling $\nubar$ from left to right.
We find that, as $\nubar$ increases, the emergence of a maximum in $\Pmp$ becomes more pronounced
indicating the emergence of a bound state between two RTPs with opposite orientation.
Thus, counterintuitively, a stronger repulsive force leads to effective attraction \cite{letter}.
The emergence of bound states, although weaker, is also found between RTPs with equal orientation, as
indicated by the two maxima in $\Ppp$. There is no bare mechanism in the Langevin \Eref{oLEq} that generates
these bound states. Their origin lies in the effective resetting at short distances that takes place through tumbling,
as one of the head-on colliding particles reorients, resulting in same-species particles
being found close to each other.
In \fref{corr_funcs_Pe} we show the same correlation functions but for increasing activity $\Pe$ from left to right
at fixed interaction coupling $\nubar$. Driving the system away from equilibrium by activity results,
similar to \fref{corr_funcs_nubar}, in the
emergence of bound states between both RTPs independently of their relative orientations.
As $\Pe$ increases, the position of the maximum in $\Pmp$, 
which defines the accumulation distance $x_{\text{A}}$ \cite{letter}, decreases.
In other words, both RTPs tend to 
accumulate at closer distance from each other on average, as shown in Fig.~3(b) of \cite{letter}.

\subsection{Overlap probability}
\begin{figure}
\includegraphics[width=0.45\textwidth]{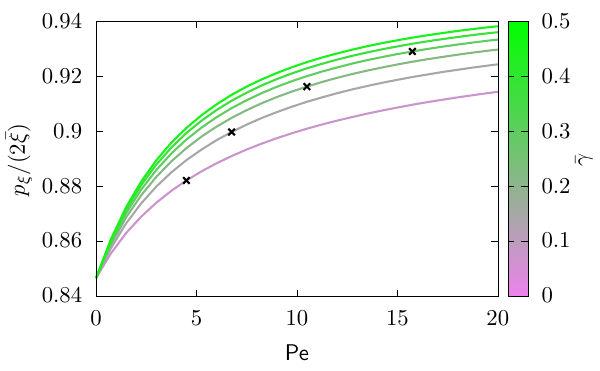}
\caption{\flabel{overlap_prob}
Overlap probability $p_\xi$ rescaled by its value in the non-interacting case, $p_\xi = 2\xi/L $,
as a function of activity $\Pe$ for varying tumbling rate $\gammabar$.
Symbols indicate the value of $\Pe$ at which the compressibility factor $S_1$ 
changes sign from $S_1<0$ (small $\Pe$, effective repulsion) to $S_1>0$ (large $\Pe$, effective attraction).
Parameters: 
$\diff = 1$,
$L = 20$,
$\nubar = 5$, and
$\xibar = 0.01$.
}
\end{figure}

Having characterised the steady state analytically, we now use the correlation functions to calculate
other quantities. An observable of interest is, for instance, the overlap probability $p_\xi$, namely the probability
that the two particles are found at a shorter distance than $\xi$,
\begin{equation}
p_\xi = \frac{1}{2} \int_{-L/2}^{L/2} \dint{x_1} \int_{x_1-\xi}^{x_1+\xi} \dint{x_2}  P(x_1,x_2) \ .
\end{equation}
Since both particles are soft, they overlap and even cross each other due to fluctuations.
The overlap probability 
rescaled by its value in the non-interacting case $p_\xi=2\xibar$, where $P(x_1,x_2)=2/L^2$,
is shown in \fref{overlap_prob}
as a function of activity $\Pe$ for varying tumbling rate $\gammabar$. 
We find that the overlap probability $p_\xi$ increases with increasing activity and increasing tumbling rate.
The underlying jamming mechanism is the cause for this increase in overlap.

\subsection{Entropy production rate}
\begin{figure}
\includegraphics[width=0.45\textwidth]{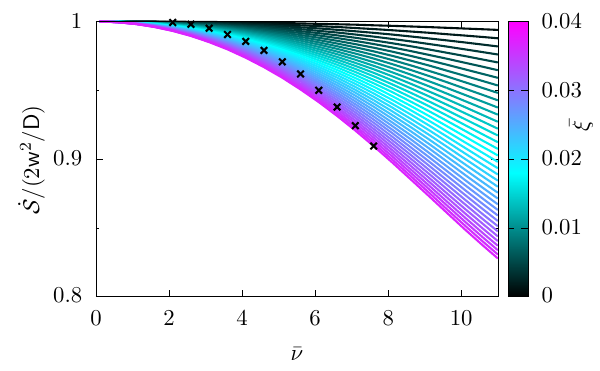}
\caption{\flabel{EPR}
Entropy production rate $\epr$
 rescaled by its value in the non-interacting case, $\epr = 2 \drift^2/\diff $,
as a function of interaction coupling $\nubar$ for varying interaction length $\xibar$.
Symbols indicate the value of $\nubar$ at which the compressibility factor $S_1$ 
changes sign from $S_1<0$ (small $\nubar$, effective repulsion) to $S_1>0$ (large $\nubar$, effective attraction).
Parameters: 
$\diff = 1$,
$L = 20$,
$\gammabar = 0.05$, and
$\Pe = 10$.
}
\end{figure}

Another observable of interest is the entropy production rate $\epr$, which quantifies the temporal irreversibility
of a system out of equilibrium \cite{PruessnerGarcia-Millan:2025,Sekimoto:2010},
\begin{align}
\elabel{eprgen}
    \epr = & \frac{\drift^2}{\diff}
    + \frac{4}{\diff}\int_{-L/2}^{L/2}\dint{x} \left( -\diff \pairPot''(x) + (\pairPot'(x))^2 \right) \Ppp(x) \nonumber\\&
+ \frac{4}{\diff}\int_{-L/2}^{L/2}\dint{x} \left( -\diff \pairPot''(x) + (\drift - \pairPot'(x))^2 \right) \Ppm(x) \ .
\end{align}
A free run-and-tumble particle produces entropy at rate $\drift^2/\diff$ \cite{CocconiETAL:2020}. Here, we use
the correlation functions $\Ppp$ and $\Pmp$ calculated analytically, \sref{corr_funcs},
to characterise how the entropy production varies as the soft repulsion $\nubar$ is 
increased.
As shown in \fref{EPR}, both an increasing repulsion as well as an increasing interaction length lead
to a decreased rate of entropy production
as particles jam more frequently, thereby effectively slowing down.

\section{Discussion and Conclusion}
\seclabel{conclusion}
In this paper we have characterised soft, interacting run-and-tumble particles
in a one-dimensional, periodic domain using a microscopic field theory.
By means of an iterative scheme, we show how the effective interaction vertices are calculated analytically,
characterising the stationary state of the two-particle system.
We further use the effective interaction vertices to calculate a number of observables that quantify
the emergence of effective interaction \cite{letter}.

The effective interaction vertices are calculated in a systematic perturbation expansion in the interaction coupling.
We include finite-size effects by accounting for discrete wavenumbers $k_j$ in the Fourier convention
and by using Matsubara frequency sums to calculate the loop corrections.
More and more terms are generated in the ensuing iterative scheme. In principle all terms can be tracked,
however, we pragmatically neglect some that we demonstrate are multiplied by an 
exponentially small coefficient provided $L\xi^{-1}$, $\gamma L^2/\diff\gg1$.

Our iterative method exploits the simple, somewhat propagator-like
form of the Yukawa potential, which makes the terms more easily tractable at every order.
From the form of the Yukawa interaction potential also stems the fact that the effective interaction vertices and 
pair correlation functions are linear combinations of the rescaled potential and its spatial derivative.
Extending our method to a general soft interaction potential and many interacting particles
remains an open question
of great interest.

\section*{Acknowledgements}
RG-M was supported in part by the European Research Council under the EU's Horizon 2020 
Programme (Grant number 740269), 
and acknowledges support from a St John's College Research Fellowship, University of Cambridge.

\begin{widetext}
\appendix
\section{Diagrammatic iteration}
\seclabel{details_iteration}
This section shows the
diagrammatic representation of how the effective interaction vertices 
$\HeteroPot_{n+1}$, $\tHomoPot_{n+1}$ and $\tHeteroPot_{n+1}$
are generated from
$\HomoPot_n$, $\HeteroPot_n$, $\tHomoPot_n$ and $\tHeteroPot_n$,
similarly to $\HomoPot_{n+1}$ in
\Eref{HomoPot_recurrence_diag},
\begin{subequations}
\elabel{PotsRecursion_diag}
\begin{align}
     \HeteroPot_{n+1}(k_j)  \corresponds&
\begin{aligned}[t]
&
\renewcommand{\LoopDiagramPot}{\HomoPot_n(k_i)}
\LoopDiagram{tAsubstrate}{}{Aactivity}{}{-k_i}{Aactivity}{Aactivity}{draw=none}{}
\ +
\renewcommand{\LoopDiagramPot}{\HomoPot_n(-k_i)}
\LoopDiagram{tAsubstrate}{}{Aactivity}{draw=none}{}{Aactivity}{Aactivity}{}{k_i}
\ +
\renewcommand{\LoopDiagramPot}{\HeteroPot_n(k_i)}
\LoopDiagram{tAsubstrate}{}{tAsubstrate}{}{-k_i}{Aactivity}{Aactivity}{draw=none}{}
\ +
\renewcommand{\LoopDiagramPot}{\tHeteroPot_n(-k_i)}
\LoopDiagram{tAsubstrate}{}{tAsubstrate}{draw=none}{}{Aactivity}{Aactivity}{}{k_i} 
\\
+&
\renewcommand{\LoopDiagramPot}{\tHeteroPot_n(k_i)}
\LoopDiagram{tAsubstrate}{}{Aactivity}{}{-k_i}{Aactivity}{tAsubstrate}{draw=none}{}
\ +
\renewcommand{\LoopDiagramPot}{\HeteroPot_n(-k_i)}
\LoopDiagram{tAsubstrate}{}{Aactivity}{draw=none}{}{Aactivity}{tAsubstrate}{}{k_i}
\ +
\renewcommand{\LoopDiagramPot}{\tHomoPot_n(k_i)}
\LoopDiagram{tAsubstrate}{}{tAsubstrate}{}{-k_i}{Aactivity}{tAsubstrate}{draw=none}{}
\ +
\renewcommand{\LoopDiagramPot}{\tHomoPot_n(-k_i)}
\LoopDiagram{tAsubstrate}{}{tAsubstrate}{draw=none}{}{Aactivity}{tAsubstrate}{}{k_i} \ ,
\end{aligned}
\elabel{HeteroPot_recurrence_diag}\\
\tHeteroPot_{n+1}(k_j)  \corresponds& 
\begin{aligned}[t]
&
\renewcommand{\LoopDiagramPot}{\HomoPot_n(k_i)}
\LoopDiagram{Aactivity}{}{Aactivity}{}{-k_i}{tAsubstrate}{Aactivity}{draw=none}{}
\ +
\renewcommand{\LoopDiagramPot}{\HomoPot_n(-k_i)}
\LoopDiagram{Aactivity}{}{Aactivity}{draw=none}{}{tAsubstrate}{Aactivity}{}{k_i}
\ +
\renewcommand{\LoopDiagramPot}{\HeteroPot_n(k_i)}
\LoopDiagram{Aactivity}{}{tAsubstrate}{}{-k_i}{tAsubstrate}{Aactivity}{draw=none}{}
\ +
\renewcommand{\LoopDiagramPot}{\tHeteroPot_n(-k_i)}
\LoopDiagram{Aactivity}{}{tAsubstrate}{draw=none}{}{tAsubstrate}{Aactivity}{}{k_i} \\
+ &
\renewcommand{\LoopDiagramPot}{\tHeteroPot_n(k_i)}
\LoopDiagram{Aactivity}{}{Aactivity}{}{-k_i}{tAsubstrate}{tAsubstrate}{draw=none}{}
\ +
\renewcommand{\LoopDiagramPot}{\HeteroPot_n(-k_i)}
\LoopDiagram{Aactivity}{}{Aactivity}{draw=none}{}{tAsubstrate}{tAsubstrate}{}{k_i}
\ +
\renewcommand{\LoopDiagramPot}{\tHomoPot_n(k_i)}
\LoopDiagram{Aactivity}{}{tAsubstrate}{}{-k_i}{tAsubstrate}{tAsubstrate}{draw=none}{}
\ +
\renewcommand{\LoopDiagramPot}{\tHomoPot_n(-k_i)}
\LoopDiagram{Aactivity}{}{tAsubstrate}{draw=none}{}{tAsubstrate}{tAsubstrate}{}{k_i} \ ,
\end{aligned}
\elabel{tHeteroPot_recurrence_diag}\\
\tHomoPot_{n+1}(k_j)  \corresponds& 
\begin{aligned}[t]
&
\renewcommand{\LoopDiagramPot}{\HomoPot_n(k_i)}
\LoopDiagram{tAsubstrate}{}{Aactivity}{}{-k_i}{tAsubstrate}{Aactivity}{draw=none}{}
\ +
\renewcommand{\LoopDiagramPot}{\HomoPot_n(-k_i)}
\LoopDiagram{tAsubstrate}{}{Aactivity}{draw=none}{}{tAsubstrate}{Aactivity}{}{k_i}
\ +
\renewcommand{\LoopDiagramPot}{\HeteroPot_n(k_i)}
\LoopDiagram{tAsubstrate}{}{tAsubstrate}{}{-k_i}{tAsubstrate}{Aactivity}{draw=none}{}
\ +
\renewcommand{\LoopDiagramPot}{\tHeteroPot_n(-k_i)}
\LoopDiagram{tAsubstrate}{}{tAsubstrate}{draw=none}{}{tAsubstrate}{Aactivity}{}{k_i} \\
+ &
\renewcommand{\LoopDiagramPot}{\tHeteroPot_n(k_i)}
\LoopDiagram{tAsubstrate}{}{Aactivity}{}{-k_i}{tAsubstrate}{tAsubstrate}{draw=none}{}
\ +
\renewcommand{\LoopDiagramPot}{\HeteroPot_n(-k_i)}
\LoopDiagram{tAsubstrate}{}{Aactivity}{draw=none}{}{tAsubstrate}{tAsubstrate}{}{k_i}
\ +
\renewcommand{\LoopDiagramPot}{\tHomoPot_n(k_i)}
\LoopDiagram{tAsubstrate}{}{tAsubstrate}{}{-k_i}{tAsubstrate}{tAsubstrate}{draw=none}{}
\ +
\renewcommand{\LoopDiagramPot}{\tHomoPot_n(-k_i)}
\LoopDiagram{tAsubstrate}{}{tAsubstrate}{draw=none}{}{tAsubstrate}{tAsubstrate}{}{k_i} \ .
\end{aligned}
\elabel{tHomoPot_recurrence_diag}
\end{align}
\end{subequations}
The analytic expressions corresponding to \Eref{PotsRecursion_diag}
are in \Eref{PotsRecursion}.

\section{Symmetries of the effective interaction vertices}
\seclabel{symmetriesPhiPsi}
    In this section we derive \Eref{Pot_n_induction_condition} by weak induction
    using \eref{pots1} as base case $n=1$ and the induction step in \eref{PotsRecursion}.
    One readily verifies that \eref{pots1} satisfies \eref{Pot_n_induction_condition}.
We first show \Eref{HomoPot_n_induction_condition}.
Changing the sign of the dummy variable $k_i$ in \Eref{tHomoPot_recurrence} by substituting
$k_i\mapsto-k_i$ effectively changes every $E$ into $G$ and vice versa, \Erefs{prop_conventions_G} and \eref{prop_conventions_E}, and leaves $D$ unchanged, \Eref{prop_conventions_D}. Further, under $k_i\mapsto-k_i$ every dashed potential becomes undashed and vice versa, so that \Eref{tHomoPot_recurrence} gives
\begin{align}
    \tHomoPot_{n+1}(k_j)=&(-k_j)\frac{1}{L}
    \sum_{
    \substack{i\in\Zset\\
    i\ne0}}
    \int\dintbar{\omega'} 
\frac{\nu \xi^{-2} (k_j+k_i)}{(k_j+k_i)^2+\xi^{-2}}DD' 
\Big\{
\begin{aligned}[t]
  &HH (\HomoPot_n'+\HomoPot_n)
+ GH (\HeteroPot_n'+\tHeteroPot_n) \\
+ & HG' (\tHeteroPot_n'+\HeteroPot_n)
+ GG' (\tHomoPot_n'+\tHomoPot_n)
\Big\} \ .
\end{aligned}
\end{align}
Evaluating $\tHomoPot_{n+1}$ at $-k_j$ and using \Eref{Pot_n_induction_condition}
we have
\begin{align}
    \tHomoPot_{n+1}(-k_j)=&k_j\frac{1}{L}
    \sum_{
    \substack{i\in\Zset\\
    i\ne0}}
    \int\dintbar{\omega'} 
\frac{\nu \xi^{-2} (-k_j+k_i)}{(-k_j+k_i)^2+\xi^{-2}}DD'
\Big\{
\begin{aligned}[t]
  &HH (\tHomoPot_n'+\tHomoPot_n)
+ GH (\HeteroPot_n'+\tHeteroPot_n) 
\\
+ &HG' (\tHeteroPot_n'+\HeteroPot_n)
+ GG' (\HomoPot_n'+\HomoPot_n)
\Big\} \ ,
\end{aligned}
\end{align}
which is equal to the right-hand side of \Eref{HomoPot_recurrence}.
It follows that 
\begin{equation}
\elabel{tHomoPot_is_HomoPot}
    \tHomoPot_{n+1}(-k_j)=\HomoPot_{n+1}(k_j) \ .
\end{equation} 
The final step to demonstrate \eref{HomoPot_n_induction_condition} from \Eref{tHomoPot_is_HomoPot}
is to show that $\tHomoPot_{n+1}(k_j)$ is even in $k_j$.
By physical reasoning this follows from the equal-species effective potential $\HomoPot(x)$ 
being even in $x$, but we choose to demonstrate this on mathematical grounds.
Changing the sign of the dummy variables in \Eref{tHomoPot_recurrence}, $\omega'\mapsto-\omega'$ and $k_i\mapsto-k_i$, effectively replaces every $D,G,E$ by $D',G',E'$ and vice versa, while leaving $H$ unchanged. Similarly, every effective vertex 
gets dashed if it appears undashed, and vice versa. It follows that
\begin{align}
\tHomoPot_{n+1}(k_j)=&(-k_j)\frac{1}{L}
    \sum_{
    \substack{i\in\Zset\\
    i\ne0}}
    \int\dintbar{\omega'} 
\frac{\nu \xi^{-2} (k_j+k_i)}{(k_j+k_i)^2+\xi^{-2}}DD' 
\Big\{
\begin{aligned}[t]
&  HH (\HomoPot_n'+\HomoPot_n)
+ E'H (\HeteroPot_n'+\tHeteroPot_n) \\
+& HE (\tHeteroPot_n'+\HeteroPot_n)
+ E'E (\tHomoPot_n'+\tHomoPot_n)
\Big\}
\end{aligned}
\end{align}
which, evaluated at $-k_j$, gives
the right-hand side of \Eref{tHomoPot_recurrence}. 
It follows that $\tHomoPot_{n+1}$ is even in $k_j$
and, together with \Eref{tHomoPot_is_HomoPot},
so is $\HomoPot_{n+1}$. In summary, we have
\begin{equation}\elabel{tHomoPot_is_HomoPot_summary}
    \HomoPot_{n+1}(-k_j)=\tHomoPot_{n+1}(k_j)=
    \tHomoPot_{n+1}(-k_j)=\HomoPot_{n+1}(k_j) 
\end{equation} 
for all $n$ and thus \Eref{HomoPot_n_induction_condition}.

As a final property of $\HomoPot(k_j)$, we show that it is real 
for $k_j\in\mathbb{R}$, which, again, follows from physical reasoning that $\HomoPot(x)$ is real and even.
To show this explicitly we use that 
the mapping $\omega'\mapsto-\omega'$ and $k_i\mapsto-k_i$, which exchanges dashes.
This operation is identical to taking the complex conjugate because every term in $G,E,H,D$ that is odd in $k_j$ is imaginary and every term that is even in $k_j$ is real. The same applies to $\omega$. As a result, for example, $D^*=D'$, and therefore $(DD')^*=DD'$
is bound to be real. Assuming that
\begin{subequations}
        \elabel{Pot_complexity_n}
        \begin{align}
            \HomoPot^*_n(k_j) =& \HomoPot_n(-k_j) \ , \elabel{HomoPot_complexity}\\
            \tHomoPot^*_n(k_j) =& \tHomoPot_n(-k_j) \ , \elabel{tHomoPot_complexity}\\
            \HeteroPot^*_n(k_j) =& \HeteroPot_n(-k_j) \ , \elabel{HeteroPot_complexity}\\
            \tHeteroPot^*_n(k_j) =& \tHeteroPot_n(-k_j) \ , \elabel{tHeteroPot_complexity}
        \end{align}
\end{subequations}
which holds for the base case \eref{pots1},
gives for \Eref{HomoPot_recurrence}
\begin{align}
\HomoPot^*_{n+1}(k_j)
=&(-k_j)\frac{1}{L}
    \sum_{
    \substack{i\in\Zset\\
    i\ne0}}
    \int\dintbar{\omega'} 
 \frac{\nu \xi^{-2} (k_j-k_i)}{(k_j-k_i)^2+\xi^{-2}}DD' 
\Big\{
\begin{aligned}[t]
  &GG'(\HomoPot_n'+\HomoPot_n)
+ HG (\HeteroPot_n'+\tHeteroPot_n) 
\\
+ &G'H (\tHeteroPot_n'+\HeteroPot_n)
+ HH (\tHomoPot_n'+\tHomoPot_n)
\Big\} \ ,
\end{aligned}
\end{align}
whose right-hand side is 
identical to \Eref{HomoPot_recurrence}. Thus
\begin{equation}\elabel{HomoPot_real}
    \HomoPot^*_{n+1}(k_j)=\HomoPot_{n+1}(k_j)
    \ ,
\end{equation}
which will be combined below with the corresponding property of $\HeteroPot_n$,
\Erefs{HeteroPot_complex_is_odd}, to confirm \Erefs{Pot_complexity_n} for all $n$.
In summary, the effective vertex $\HomoPot_n(k_j)=\tHomoPot_n(k_j)$ is even and real.

To show \Eref{HeteroPot_n_induction_condition}, we repeat a similar argument for $\HeteroPot_n$,
which is slightly more intricate, because $\HeteroPot_n$ contains a real and an imaginary part.
First, replacing $k_i\mapsto-k_i$ in \Eref{tHeteroPot_recurrence} and evaluating at $-k_j$ gives
\begin{align}
\tHeteroPot_{n+1}(-k_j)=&k_j\frac{1}{L}
    \sum_{
    \substack{i\in\Zset\\
    i\ne0}}
    \int\dintbar{\omega'} 
\frac{\nu \xi^{-2} (-k_j+k_i)}{(-k_j+k_i)^2+\xi^{-2}}DD' 
\Big\{
\begin{aligned}[t]
&  EH (\HomoPot_n'+\HomoPot_n)
+ HH (\HeteroPot_n'+\tHeteroPot_n) 
\\
+ & EG' (\tHeteroPot_n'+\HeteroPot_n)
+ HG' (\tHomoPot_n'+\tHomoPot_n)
\Big\} \ ,
\end{aligned}
\end{align}
which is equal to $\HeteroPot_{n+1}(k_j)$ in \eref{HeteroPot_recurrence}, provided \Eref{Pot_n_induction_condition}. 
It follows that,
\begin{equation}\elabel{tHeteroPot_is_HeteroPot}
    \tHeteroPot_{n+1}(-k_j)=\HeteroPot_{n+1}(k_j) \ ,
\end{equation}
confirming \Eref{HeteroPot_n_induction_condition} for all $n$.
The complex conjugate of $\HeteroPot_{n+1}$ in \eref{HeteroPot_recurrence} is, 
assuming \Eref{Pot_complexity_n},
\begin{align}
    \HeteroPot^*_{n+1}(k_j)=&(-k_j)\frac{1}{L}
    \sum_{
    \substack{i\in\Zset\\
    i\ne0}}
    \int\dintbar{\omega'} 
 \frac{\nu \xi^{-2} (k_j-k_i)}{(k_j-k_i)^2+\xi^{-2}}D'D 
\Big\{
\begin{aligned}[t]
&  HG (\HomoPot_n'+\HomoPot_n)
+ E'G(\HeteroPot_n'+\tHeteroPot_n) \\
+ & HH (\tHeteroPot_n'+\HeteroPot_n)
+ E'H (\tHomoPot_n'+\tHomoPot_n)
\Big\} \ ,
\end{aligned}
\end{align}
which is exactly \Eref{tHeteroPot_recurrence}. Therefore,
\begin{equation}\elabel{HeteroPot_complex}
    \HeteroPot^*_{n+1}(k_j)=\tHeteroPot_{n+1}(k_j)
\end{equation}
and combined with \Eref{tHeteroPot_is_HeteroPot}, we have
\begin{subequations}
\elabel{HeteroPot_complex_is_odd}
\begin{align}
    \HeteroPot^*_{n+1}(k_j)= & \HeteroPot_{n+1}(-k_j) \elabel{HeteroPot_complex_is_odd_a}\\
    \tHeteroPot^*_{n+1}(k_j)= & \tHeteroPot_{n+1}(-k_j)
    \ ,
\end{align}
\end{subequations}
together with \Eref{HomoPot_real}
confirming \Erefs{Pot_complexity_n}
for arbitrary $n$
and confirming the physical picture that $\HomoPot(x)$ and $\HeteroPot(x)$ are real effective pair-potentials.

\section{Evaluation of loop integrals over $\omega$}
The loop integrals over $\omega$ in \Eref{PQR_full_integral} are evaluated
using \Erefs{def_a12} and \eref{ints_omega}.
We use the following parametrisation so as to systematically define
the iterative scheme.
For $P_{n+1}$ in \Eref{P_nP1_full_integral}, we need,
\begin{subequations}
\elabel{def_f}
\begin{align}\elabel{def_fP}
    \int \dintbar{\omega'} 2 DD' (GG'+HH) k_i^2 = & f_P(k_i) \nonumber\\
    = & \varphi_{P0} + \frac{\varphi_{P1}}{k_i^2+a_1^2} + \frac{\varphi_{P2}}{k_i^2+a_2^2} \nonumber\\ 
    = & \frac{1}{\diff} 
    - \frac{\gamma\drift^2}{\diff^2} \frac{1}{\gamma\diff+\drift^2} \frac{1}{k_i^2+a_1^2}
    - \frac{1}{\diff} \frac{\gamma^2}{\gamma\diff+\drift^2} \frac{1}{k_i^2+a_2^2} \ ,\\
    \elabel{def_fQ}
    \int \dintbar{\omega'} 2 DD' (HG'+GH) k_i^2 =&  f_Q(k_i) \nonumber\\
    =&\varphi_{Q0} + \frac{\varphi_{Q1}}{k_i^2+a_1^2} + \frac{\varphi_{Q2}}{k_i^2+a_2^2}  \nonumber\\ 
    =&\frac{\gamma}{\diff^2} \frac{1}{k_i^2+a_2^2} \ , \\
    \int \dintbar{\omega'} 2 DD' (HG'-GH) \imag k_i =&  f_R(k_i) \nonumber\\
    =&\varphi_{R0} + \frac{\varphi_{R1}}{k_i^2+a_1^2} + \frac{\varphi_{R2}}{k_i^2+a_2^2} \nonumber\\ 
    =&-\frac{\drift\gamma}{\diff}\frac{1}{\gamma\diff+\drift^2}\frac{1}{k_i^2+a_1^2}
     +\frac{\drift\gamma}{\diff}\frac{1}{\gamma\diff+\drift^2}\frac{1}{k_i^2+a_2^2} \ .
\end{align}
\end{subequations}

Similarly, for $Q_{n+1}$ in \Eref{Q_nP1_full_integral} we need,
\begin{subequations}
\elabel{def_g}
\begin{align}
    \elabel{def_gP}
    \int \dintbar{\omega'} DD' (HG'+GH+EH+HE')k_i^2=&g_P(k_i) \nonumber\\
    =&\gamma_{P0} + \frac{\gamma_{P1}}{k_i^2+a_1^2} + \frac{\gamma_{P2}}{k_i^2+a_2^2} \nonumber\\
    =&\frac{\gamma}{\diff^2} \frac{1}{k_i^2+a_2^2} \ ,\\
    \elabel{def_gQ}
    \int \dintbar{\omega'} DD' (EG'+HH+HH+GE')k_i^2=&g_Q(k_i) \nonumber\\
    =&\gamma_{Q0} + \frac{\gamma_{Q1}}{k_i^2+a_1^2} + \frac{\gamma_{Q2}}{k_i^2+a_2^2}\nonumber\\
    =&\frac{1}{\diff}-\frac{\gamma\diff+\drift^2}{\diff^3}\frac{1}{k_i^2+a_2^2} \ ,\\
    \elabel{def_gR}
    \int \dintbar{\omega'} DD' (EG'+HH-HH-GE')\imag k_i=&g_R(k_i) \nonumber\\
    =&\gamma_{R0} + \frac{\gamma_{R1}}{k_i^2+a_1^2} + \frac{\gamma_{R2}}{k_i^2+a_2^2}\nonumber\\
    =&\frac{-\drift}{\diff^2} \frac{1}{k_i^2+a_2^2} \ .
\end{align}
\end{subequations}

Further, for $R_{n+1}$ in \Eref{R_nP1_full_integral} we need,
\begin{subequations}
\elabel{def_h}
\begin{align}
    \elabel{def_hP}
    \int \dintbar{\omega'} DD' (HG'-GH+EH-HE')k_i^2=& \imag k_i h_P(k_i) \nonumber\\
    =&\imag k_i\bigg(\eta_{P0} + \frac{\eta_{P1}}{k_i^2+a_1^2} + \frac{\eta_{P2}}{k_i^2+a_2^2}\bigg)\nonumber\\
    =&\imag k_i \bigg(
    \frac{\gamma\drift}{\diff(\gamma\diff+\drift^2)}\frac{1}{k_i^2+a_1^2}
    - 
    \frac{\gamma\drift}{\diff(\gamma\diff+\drift^2)}\frac{1}{k_i^2+a_2^2}
    \bigg) \ ,\\
    \int \dintbar{\omega'} DD' (EG'-HH+HH-GE')k_i^2=& \imag k_i h_Q(k_i) \nonumber\\
    =&\imag k_i\bigg(\eta_{Q0} + \frac{\eta_{Q1}}{k_i^2+a_1^2} + \frac{\eta_{Q2}}{k_i^2+a_2^2}\bigg)\nonumber\\
    =&\imag k_i\bigg(\frac{\drift}{\diff^2} \frac{1}{k_i^2+a_2^2}\bigg) \ , \\
    \int \dintbar{\omega'} DD' (EG'-HH-HH+GE')\imag k_i=& \imag k_i h_R(k_i) \nonumber\\
    =&\imag k_i\bigg(\eta_{R0} + \frac{\eta_{R1}}{k_i^2+a_1^2} + \frac{\eta_{R2}}{k_i^2+a_2^2}\bigg)\nonumber\\
    =&\imag k_i\bigg(
    \frac{\gamma}{\gamma \diff + \drift^2} \frac{1}{k_i^2+a_1^2} 
    +
    \frac{\drift^2/\diff}{\gamma \diff + \drift^2} \frac{1}{k_i^2+a_2^2} 
    \bigg) \ .
\end{align}
\end{subequations}

\section{Evaluation of loop sums over $k_j$ as infinite series}
\seclabel{app_sums}

\begin{figure}
\includegraphics[width=0.5\textwidth]{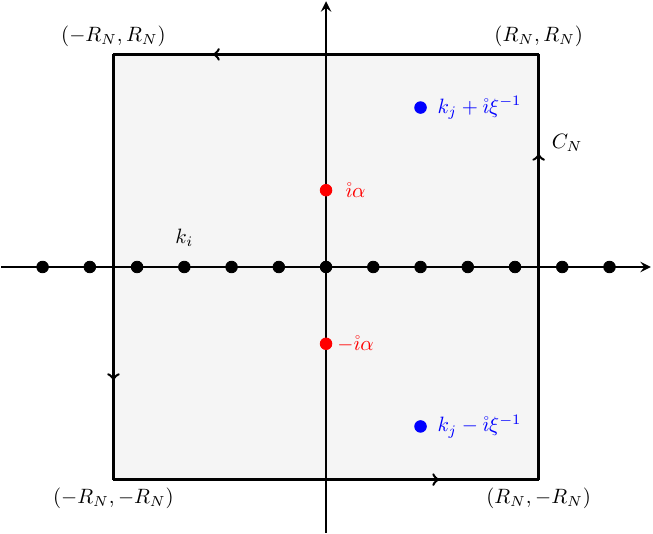}
\caption{\flabel{contour_CN} Square contour $C_N$ for $N=4$ and simple poles of $g(z)=\cot{\left({Lz}/{2} \right)} f(z)$,
 \Eref{cont_int_def}.
The poles of $\cot{\left({Lz}/{2} \right) }$ are $z=k_i$ (black), and the poles of 
$f(z)$ are $z=\pm\imag\alpha$ (red) and $z=k_j\pm\imag\xi^{-1}$ (blue) with $j=2$, 
\Erefs{def_f_sum0} and \eref{def_f_sum1}.}
\end{figure}

The infinite series in \Erefs{ksum} have the form
\begin{equation}
\frac{1}{L}
    \sum_{
    \substack{i\in\Zset\\
    i\ne0}}
    f(k_i)
    = \frac{1}{L} \left(
    \sum_{i=-\infty}^{\infty}
    f(k_i)
    -f(0)\right) \ ,
\end{equation}
with wavenumber $k_i=2\pi i /L$, and
are evaluated by applying 
Cauchy's residue theorem \cite{SpiegelETAL:2009,MarsdenHoffman:1973}.
We consider the contour integral 
$\oint_{C_N}\dint{z} g(z)$ of
\begin{equation}
\elabel{cont_int_def}
    g(z) = \cot{\left(\frac{Lz}{2} \right) } f(z)
\end{equation}
where the contour $C_N$ is a square defined by the four vertices $(\pm R_N, \pm R_N)$
with $R_N = 2\pi(N+1/2)/L$
and $N\in\mathbb{N}$
and $f$ is the summand in \Erefs{ksum}, \fref{contour_CN}.
As $|\cot(Lz/2)|<2$ along the contour $C_N$ with $N\geq1$ and both summands in \Erefs{ksum} obey
$|f(z)|\leq1/z^2$, the integrand along the contour of length $\sim N$ is suitably bounded by $\sim N^{-2}$,
namely $   \left| g(z) \right| < {2}/{|z|^2} $ for $z\in C_N$, so that
\begin{equation}
\elabel{contour_int_Ninf}
   \lim_{N\to\infty} \oint_{C_N}\dint{z} g(z) = 0 \ .
\end{equation}
This integrand has simple poles at the singularities of $\cot{\left({Lz}/{2} \right) }$
and $f(z)$ inside $C_N$, which are $z=k_i$ for all $i\in\mathbb{Z}$ with $k_i\in\mathbb{R}$
and $\{z_1,z_2,\ldots\}$ with $z_i\notin\mathbb{R}$ respectively, to be determined below.
According to Cauchy's residue theorem \cite{SpiegelETAL:2009,MarsdenHoffman:1973}, 
this contour integral equals the sum of the residues 
at the poles enclosed by the contour,
\begin{equation}
\elabel{contour_int_Nfin}
    \oint_{C_N}\dint{z} \cot{\left(\frac{Lz}{2} \right) } f(z)
    = 2\pi\imag \left[\sum_{i=-N}^N f(k_i)\Resm{z=k_i}{\cot(z)}+ \sum_{i} \cot(z_i)\Resm{z=z_i}{f(z)}\right] \ .
\end{equation}
As the residues of the poles of $\cot{\left({Lz}/{2} \right) }$ at $z=k_i$ for $i\in\mathbb{Z}$ are $2/L$,
combining \Erefs{contour_int_Ninf} and \eref{contour_int_Nfin} implies
\begin{equation}
 \frac{1}{L} \sum_{i=-\infty}^\infty f(k_i) = - \frac{1}{2}\sum_{i} \cot(z_i)\Resm{z=z_i}{f(z)}\ ,
\end{equation}
thus reducing an infinite sum to that over a small, finite number of terms.
To evaluate the series in \Eref{k0sum}, we choose
\begin{equation}
\elabel{def_f_sum0}
 f(z) =   
    \frac{k_j-z}{(k_j-z)^2+\xi^{-2}} \frac{1}{z^2+\alpha^2} \ ,
\end{equation}
which has simple poles of $f(z)$ are $z= \pm\imag\alpha$ and $z= k_j \pm\imag\xi^{-1}$, \fref{contour_CN}, with 
the following residues,
\begin{subequations}
\begin{align}
\Resm{z=\imag\alpha}{f(z)}=&\frac{1}{2\imag\alpha}  \frac{k_j-\imag\alpha}{(k_j-\imag\alpha)^2+\xi^{-2}} \ , \\
\Resm{z=-\imag\alpha}{f(z)}=& -\frac{1}{2\imag\alpha}    \frac{k_j+\imag\alpha}{(k_j+\imag\alpha)^2+\xi^{-2}} \ , \\
\Resm{z=k_j+\imag\xi^{-1}}{f(z)} =& -\frac{1}{2} 
\frac{1}{(k_j+\imag\xi^{-1})^2+\alpha^2}\ , \\
\Resm{z=k_j-\imag\xi^{-1}}{f(z)} =& -\frac{1}{2} 
\frac{1}{(k_j-\imag\xi^{-1})^2+\alpha^2} \ .
\end{align}
\end{subequations}
Using the identity
\begin{equation}
   \cot{\left(\frac{L(k_j+\imag\alpha)}{2} \right) } = \imag\frac{e^{-L\alpha}+1}{e^{-L\alpha}-1} \ ,
\end{equation}
we arrive at
\begin{subequations}
\begin{align}
    \frac{1}{L} \sum_{i=-\infty}^\infty f(k_i) 
    =
& -\frac{\imag}{2}
\begin{aligned}[t]
& \left(\frac{1}{2\alpha\imag}\frac{\e{-L\alpha}+1}{\e{-L\alpha}-1}\frac{k_j-\imag\alpha}{(k_j-\imag\alpha)^2+\xi^{-2}}
-\frac{1}{2\alpha\imag}\frac{\e{L\alpha}+1}{\e{L\alpha}-1}\frac{k_j+\imag\alpha}{(k_j+\imag\alpha)^2+\xi^{-2}}
\right.
\\ 
    & \left.
    -\frac{1}{2}\frac{\e{-L\xi^{-1}}+1}{\e{-L\xi^{-1}}-1}\frac{1}{(k_j+\imag\xi^{-1})^2+\alpha^2}
    -\frac{1}{2}\frac{\e{L/\xi}+1}{\e{L/\xi}-1}\frac{1}{(k_j-\imag\xi^{-1})^2+\alpha^2} \right)
\end{aligned} \\
= &\frac{k_j}{2\alpha}
     \left(
 \frac{\Aa(\alpha)}{k_j^2+(\alpha+\xi^{-1})^2}
+
 \frac{\Ba(\alpha)}{k_j^2+(\alpha-\xi^{-1})^2}
 \right) \ ,
\end{align}
\end{subequations}
with coefficients $A(\alpha)$ and $B(\alpha)$ defined in \Erefs{def_AB}.
We obtain \Eref{k0sum} by subtracting the zero-mode, 
$f(0) = {k_j}/\left(\alpha^2{(k_j^2+\xi^{-2})} \right)$, which produces its final term algebraic in $L$.
The evaluation of the infinite series in \Eref{k1sum} follows the same steps as above, giving
\begin{subequations}
\elabel{def_f_sum1}
\begin{align}
\frac{1}{L} \sum_{i=-\infty}^\infty f(k_i) = &    \frac{1}{L}
    \sum_{i=-\infty}^\infty
    \frac{k_j-k_i}{(k_j-k_i)^2+\xi^{-2}} \frac{k_i}{k_i^2+\alpha^2} \\
     = & -\frac{1}{2}
 {\left(
   \frac{\Ab(\alpha) (\alpha+\xi^{-1})}{k_j^2+(\alpha+\xi^{-1})^2} 
+ \frac{\Bb(\alpha) (\alpha-\xi^{-1})}{k_j^2+(\alpha-\xi^{-1})^2} 
 \right)}
    \ .
\end{align}
\end{subequations}
In this case, removing the zero-mode leaves the sum unchanged because $f(0)=0$.

\section{Iteration of the functions $P$, $Q$ and $R$}
\label{iteration_spelled_out}
The sets of poles of each type, $p^{(0)}_{n,m}$, $p^{(1)}_{n,m}$ and $p^{(2)}_{n,m}$, listed in \Eref{pole_types} 
have indices
$m\in[1,M_n^{(0)}]$ and $m\in[-M_n^{(i)},M_n^{(i)}]$, with $M_n^{(0)}=n$ and $M_n^{(i)}=n-1$ for $i\in\{1,2\}$.
Henceforth, denoting any pole in the set of $M_n = M_n^{(0)} + 2 M_n^{(1)} + 2 M_n^{(2)} + 2$ 
poles as $p_{n,m}$, we use the shorthand notation
\begin{align}
\sum_{m=1}^{M_n}f(p_{n,m}) = 
  \sum_{\mtilde=1}^{\Mtilde_n}f(p_{n,\mtilde}^{(0)})
      +
    \sum_{m=-M^{(1)}_n}^{M^{(1)}_n}f(p^{(1)}_{n,m})
    +
    \sum_{m=-M^{(2)}_n}^{M^{(2)}_n}f(p^{(2)}_{n,m})
\end{align}
for arbitrary $f$.
For better readability, we introduce the bracket notation
\begin{subequations}
\elabel{bracket_notation}
\begin{align}
[\varphi]_{n,m} &= 
 \begin{aligned}[t]
 \bigg[
 &\Big(\varphi_{P0} + \frac{\varphi_{P1}}{a_1^2-p_{n,m}^2} + \frac{\varphi_{P2}}{a_2^2-p_{n,m}^2}\Big)\pi_{n,m}\\
 +
 &\Big(\varphi_{Q0} + \frac{\varphi_{Q1}}{a_1^2-p_{n,m}^2} + \frac{\varphi_{Q2}}{a_2^2-p_{n,m}^2}\Big)\zeta_{n,m}\\
 +
 &\Big(\varphi_{R0} + \frac{\varphi_{R1}}{a_1^2-p_{n,m}^2} + \frac{\varphi_{R2}}{a_2^2-p_{n,m}^2}\Big)\rho_{n,m}
 \bigg] \ ,
 \end{aligned}
\elabel{square_bracket_notation}
 \\
    |\varphi_1|_n &= 
    \sum_{m=1}^{M_n} \bigg[
 \frac{\varphi_{P1}}{a_1^2-p_{n,m}^2} \pi_{n,m}
 +
 \frac{\varphi_{Q1}}{a_1^2-p_{n,m}^2} \zeta_{n,m}
 +
 \frac{\varphi_{R1}}{a_1^2-p_{n,m}^2} \rho_{n,m}
 \bigg] \ ,\\
    |\varphi_2|_n &= 
    \sum_{m=1}^{M_n} \bigg[
 \frac{\varphi_{P2}}{a_2^2-p_{n,m}^2} \pi_{n,m}
 +
 \frac{\varphi_{Q2}}{a_2^2-p_{n,m}^2} \zeta_{n,m}
 +
 \frac{\varphi_{R2}}{a_2^2-p_{n,m}^2} \rho_{n,m}
 \bigg]
  \ .
\end{align}
\end{subequations}

Using the parametrisation in \Erefs{PQR_reparam} and \eref{bracket_notation},
which assumes simple poles $\pm\imag p^{(i)}_{n,m}$,
and the identity \eref{partial_fractions},
the next order $P_{n+1}(k_j)$ of $P_n$ in \Eref{P_nP1_k_integral} reads
\begin{align} 
P_{n+1}(k_j)k_j^2 = (-k_j)
    \frac{1}{L}
\sum_{\substack{i\in\Zset\\
i\ne0}}
(k_j-k_i) \frac{(\nu\xi^{-2})^{n+1}}{(k_j-k_i)^2+\xi^{-2}} 
\Bigg\{
 \sum_{m=1}^{M_n}  \frac{1}{k_i^2+p_{n,m}^2}
 [\varphi]_{n,m}
 - \frac{1}{k_i^2+a_1^2} 
 |\varphi_1|_n
 - \frac{1}{k_i^2+a_2^2}
 |\varphi_2|_n
 \Bigg\} \ .
\elabel{P_nP1_k_sum_rearranged}
\end{align}
Using now the Matsubara sum in \Eref{k0sum},
the loop sum over $k_i$ produces, 
\begin{subequations}
\elabel{P_nP1_k_sum_done}
\begin{align}
P_{n+1}(k_j)=-
(\nu\xi^{-2})^{n+1} 
\Bigg\{
\elabel{P_nP1_k_sum_done_purple2}
 &
 - 
 |\varphi_1|_n
 \frac{1}{2a_1} 
 {\left(
 \frac{\Aa(a_1)}{k_j^2+(a_1+\xi^{-1})^2} 
+
 \frac{\Ba(a_1)}{k_j^2+(a_1-\xi^{-1})^2} 
-
 \frac{2}{L a_1}\frac{1}{k_j^2+\xi^{-2}}
 \right)} \\
\elabel{P_nP1_k_sum_done_purple3}
&
 - 
 |\varphi_2|_n 
 \frac{1}{2a_2}
 {\left(
 \frac{\Aa(a_2)}{k_j^2+(a_2+\xi^{-1})^2} 
+
 \frac{\Ba(a_2)}{k_j^2+(a_2-\xi^{-1})^2} 
-
 \frac{2}{L a_2}\frac{1}{k_j^2+\xi^{-2}}
 \right)}
\\
& +
 \sum_{m=1}^{M_n} 
  [\varphi]_{n,m}
 \frac{1}{2p_{n,m}}
 {\left(
 \frac{\Aa(p_{n,m})}{k_j^2+(p_{n,m}+\xi^{-1})^2}
+
 \frac{\Ba(p_{n,m})}{k_j^2+(p_{n,m}-\xi^{-1})^2}
-
 \frac{2}{L p_{n,m}}\frac{1}{k_j^2+\xi^{-2}}
 \right)}
 \elabel{P_nP1_k_sum_done_purple1}
 \Bigg\} \ .
\end{align}
\end{subequations}
The process for $Q_n$, \Eref{def_QR_n}, is similar. 
Using \Erefs{PQR_reparam} and \eref{partial_fractions},
$Q_{n+1}(k_j)$ in \Eref{Q_nP1_k_integral} reads
\begin{align}\elabel{Q_nP1_k_sum_rearranged}
Q_{n+1}k_j^2&=(-k_j)
    \frac{1}{L}
\sum_{\substack{i\in\Zset\\
i\ne0}}
(k_j-k_i) \frac{(\nu\xi^{-2})^{n+1}}{(k_j-k_i)^2+\xi^{-2}} 
\Bigg\{
 \sum_{m=1}^{M_n} [\gamma]_{n,m}
 \frac{1}{k_i^2+p_{n,m}^2}
- \frac{1}{k_i^2+a_1^2} 
 |\gamma_1|_n 
- \frac{1}{k_i^2+a_2^2} 
 |\gamma_2|_n  \Bigg\}\ ,
\end{align}
where the notation $[\gamma]_{n,m}$, $|\gamma_1|_n$ and $|\gamma_2|_n$
follow the same pattern as the notation in \Eref{bracket_notation}
with $\varphi$-coefficients replaced by $\gamma$-coefficients.
Using the sum in \Eref{k0sum},
\begin{subequations}
\elabel{Q_nP1_k_sum_done}
\begin{align}
Q_{n+1}(k_j)=-
(\nu\xi^{-2})^{n+1}
\Bigg\{&
\elabel{Q_nP1_k_sum_done_purple2}
 -  |\gamma_1|_n
 \frac{1}{2a_1} 
 {\left(
 \frac{\Aa(a_1)}{k_j^2+(a_1+\xi^{-1})^2} 
+
 \frac{\Ba(a_1)}{k_j^2+(a_1-\xi^{-1})^2} 
-
 \frac{2}{L a_1}\frac{1}{k_j^2+\xi^{-2}}
 \right)}
\\
\elabel{Q_nP1_k_sum_done_purple3}
& 
-  |\gamma_2|_n
\frac{1}{2a_2}
 {\left(
 \frac{\Aa(a_2)}{k_j^2+(a_2+\xi^{-1})^2} 
+
 \frac{\Ba(a_2)}{k_j^2+(a_2-\xi^{-1})^2} 
-
 \frac{2}{L a_2}\frac{1}{k_j^2+\xi^{-2}}
 \right)}
 \\
&+ \sum_{m=1}^{M_n} 
[\gamma]_{n,m}
\elabel{Q_nP1_k_sum_done_purple1}
 \frac{1}{2p_{n,m}}
 {\left(
 \frac{\Aa(p_{n,m})}{k_j^2+(p_{n,m}+\xi^{-1})^2}
+
 \frac{\Ba(p_{n,m})}{k_j^2+(p_{n,m}-\xi^{-1})^2}
-
 \frac{2}{L p_{n,m}}\frac{1}{k_j^2+\xi^{-2}}
 \right)}
 \Bigg\} \ .
\end{align}
\end{subequations}
Finally, we calculate the loop sum in \Eref{R_nP1_k_integral} using
\Eref{k1sum}, which gives
\begin{subequations}
\elabel{R_nP1_k_sum_done}
\begin{align}
R_{n+1}(k_j)=
(\nu\xi^{-2})^{n+1}
\Bigg\{
\elabel{R_nP1_k_sum_done_violet2}
& 
- |\eta|_1
 \frac{1}{2}
 {\left(
  \Ab(a_1) \frac{a_1+\xi^{-1}}{k_j^2+(a_1+\xi^{-1})^2} 
+ \Bb(a_1) \frac{a_1-\xi^{-1}}{k_j^2+(a_1-\xi^{-1})^2} 
 \right)}\\
\elabel{R_nP1_k_sum_done_violet3}
& 
- |\eta|_2
 \frac{1}{2}
 {\left(
  \Ab(a_2) \frac{a_2+\xi^{-1}}{k_j^2+(a_2+\xi^{-1})^2} 
+ \Bb(a_2) \frac{a_2-\xi^{-1}}{k_j^2+(a_2-\xi^{-1})^2} 
 \right)}
 \\
\elabel{R_nP1_k_sum_done_violet1}
&+ \sum_{m=1}^{M_n} 
[\eta]_{n,m}
 \frac{1}{2}
 {\left(
  \Ab(p_{n,m}) \frac{p_{n,m}+\xi^{-1}}{k_j^2+(p_{n,m}+\xi^{-1})^2} 
+ \Bb(p_{n,m}) \frac{p_{n,m}-\xi^{-1}}{k_j^2+(p_{n,m}-\xi^{-1})^2} 
 \right)}
 \Bigg\} \ ,
\end{align}
\end{subequations}
where the notation $[\eta]_{n,m}$, $|\eta_1|_n$ and $|\eta_2|_n$
is obtained from \Eref{bracket_notation}
by replacing $\varphi$-coefficients by $\eta$-coefficients.

\section{Iteration of the coefficients $\pi$,  $\zeta$ and $\rho$}
\label{iteration_potentials_pi_zeta_rho}
Following an algorithmic approach, in this section we write explicitly
how the residue of each simple pole at order $n$ contributes additively to a particular amplitude and thus
the residues of other simple poles at 
order $n+1$ in the perturbation theory.
This is done by identifying the coefficients in \Erefs{P_nP1_k_sum_done}, \eref{Q_nP1_k_sum_done}
and \eref{R_nP1_k_sum_done} as contributions to $\pi_{n+1,m}$,  $\zeta_{n+1,m}$ and $\rho_{n+1,m}$
for suitable $m$ according to the parametrisation in \Eref{PQR_reparam}.
As a matter of pragmatism, we ignore all $a_i$-type poles at $m\leq0$, \fref{fig_poles} (right),
as they are exponentially suppressed.
We first introduce the following bracket notation, similar to \Eref{square_bracket_notation},
\begin{equation}
\begin{aligned}[t]
[\varphi]^{(i)}_{n,m}=
 \bigg[
 &\Big(\varphi_{P0} + \frac{\varphi_{P1}}{a_1^2-(p^{(i)}_{n,m})^2} + \frac{\varphi_{P2}}{a_2^2-(p^{(i)}_{n,m})^2}\Big)\pi^{(i)}_{n,m}\\
 +
 &\Big(\varphi_{Q0} + \frac{\varphi_{Q1}}{a_1^2-(p^{(i)}_{n,m})^2} + \frac{\varphi_{Q2}}{a_2^2-(p^{(i)}_{n,m})^2}\Big)\zeta^{(i)}_{n,m}\\
 +
 &\Big(\varphi_{R0} + \frac{\varphi_{R1}}{a_1^2-(p^{(i)}_{n,m})^2} + \frac{\varphi_{R2}}{a_2^2-(p^{(i)}_{n,m})^2}\Big)\rho^{(i)}_{n,m}
 \bigg]\ .
 \end{aligned}
\end{equation}

\subsection{Contributions to $\pi_{n+1,m}$}

Contributions to $\pi_{n+1,m}$ are read off from \Eref{P_nP1_k_sum_done} as follows. 
For $\xi$-type poles $p^{(0)}_{n,m}$, the $\Aa$-term in 
\Eref{P_nP1_k_sum_done_purple1}
produces an up-shifted pole of $\xi$-type, the $\Ba$-term a down-shifted $\xi$-type, but only if $m>1$. 
Similarly, for $a_1$-type poles $p^{(1)}_{n,m}$, the $\Aa$-term produces an up-shifted $a_1$-type and the $\Ba$-term a downshifted $a_1$-type without exception. 
The same happens for $a_2$-type poles $p^{(2)}_{n,m}$. 
Within 
\Eref{P_nP1_k_sum_done_purple1}, the last term always contributes a $\xi$-type pole $p^{(0)}_{n+1,1}$. 
In summary, from 
\Eref{P_nP1_k_sum_done_purple1} we deduce the following contributions to $\pi_{n+1,m}$,
\begin{subequations}
\elabel{P_nP1_k_sum_done_purple1_all}
\begin{align}
  -
 \frac{\Aa(p_{n,\mtilde}^{(0)})}{2p_{n,\mtilde}^{(0)}}
 [\varphi]_{n,\mtilde}^{(0)}
&&& 
\mapsto 
\begin{aligned}[t]
&\pitilde_{n+1,\mtilde+1}
&\text{for }\mtilde=1,\ldots,\Mtilde_n
\end{aligned}\\
 -
 \frac{\Ba(p_{n,\mtilde}^{(0)})}{2p_{n,\mtilde}^{(0)}}
 [\varphi]_{n,\mtilde}^{(0)} 
&&& 
\mapsto 
\begin{aligned}[t]
&\pitilde_{n+1,\mtilde-1}
&\text{for }\mtilde=2,\ldots,\Mtilde_n 
\end{aligned}\\
 -
 \frac{\Aa(p^{(1)}_{n,m})}{2p^{(1)}_{n,m}}
 [\varphi]_{n,m}^{(1)}
&&& 
\mapsto 
\begin{aligned}[t]
&\pi^{(1)}_{n+1,m+1}
&\text{for }m=-M^{(1)}_n,\ldots,M^{(1)}_n
\end{aligned}\\
 -
 \frac{\Ba(p^{(1)}_{n,m})}{2p^{(1)}_{n,m}}
 [\varphi]_{n,m}^{(1)}
&&& 
\mapsto 
\begin{aligned}[t]
&\pi^{(1)}_{n+1,m-1}
&\text{for }m=-M^{(1)}_n,\ldots,M^{(1)}_n
\end{aligned}\\
 -
 \frac{\Aa(p^{(2)}_{n,m})}{2p^{(2)}_{n,m}}
 [\varphi]_{n,m}^{(2)}
&&& 
\mapsto
\begin{aligned}[t]
& \pi^{(2)}_{n+1,m+1}
&\text{for }m=-M^{(2)}_n,\ldots,M^{(2)}_n
\end{aligned}\\
 -
 \frac{\Ba(p^{(2)}_{n,m})}{2p^{(2)}_{n,m}}
 [\varphi]_{n,m}^{(2)}
&&& 
\mapsto 
\begin{aligned}[t]
&\pi^{(2)}_{n+1,m-1}
&\text{for }m=-M^{(2)}_n,\ldots,M^{(2)}_n
\end{aligned}\\
 \frac{1}{L}
 \sum_{m=1}^{M_n} 
 \frac{1}{(p_{n,m})^2}
 [\varphi]_{n,m}
&&& 
\mapsto \pitilde_{n+1,1} \ .
\end{align}
\end{subequations}
Contributions from  \Eref{P_nP1_k_sum_done_purple2} are to coefficients 
 $\pi^{(1)}_{n+1,\pm1}$ and $\pi^{(0)}_{n+1,1}$,
\begin{subequations}
\elabel{P_nP1_k_sum_done_purple2_all}
\begin{align}
\elabel{P_nP1_k_sum_done_purple2_113a}
 \frac{\Aa(a_1)}{2a_1}
 |\varphi_1|_n
&&&
\mapsto \pi^{(1)}_{n+1,1}\\
\elabel{P_nP1_k_sum_done_purple2_113b}
 \frac{\Ba(a_1)}{2a_1}
 |\varphi_1|_n
&&&
\mapsto \pi^{(1)}_{n+1,-1}\\
 -\frac{1}{La_1^2}
 |\varphi_1|_n
&&&
\mapsto \pitilde_{n+1,1} \ ,
\end{align}
\end{subequations}
and, similarly, contributions from \Eref{P_nP1_k_sum_done_purple3} are to coefficients 
 $\pi^{(2)}_{n+1,\pm1}$ and $\pi^{(0)}_{n+1,1}$,
\begin{subequations}
\elabel{P_nP1_k_sum_done_purple3_all}
\begin{align}
 \frac{\Aa(a_2)}{2a_2}
 |\varphi_2|_n
&&&
\mapsto \pi^{(2)}_{n+1,1}\\
 \frac{\Ba(a_2)}{2a_2}
 |\varphi_2|_n
&&&
\mapsto \pi^{(2)}_{n+1,-1}\\
 -\frac{1}{La_2^2}
 |\varphi_2|_n
&&&
\mapsto \pitilde_{n+1,1} \ .
\end{align}
\end{subequations}

\subsection{Contributions to $\zeta_{n+1,m}$}
The result contributions to $\zeta_{n+1,m}$ from \Eref{Q_nP1_k_sum_done}
and follow the same pattern as \Erefs{P_nP1_k_sum_done_purple1_all}, \eref{P_nP1_k_sum_done_purple2_all} and \eref{P_nP1_k_sum_done_purple3_all} with 
$[\varphi]$, $|\varphi_1|$, $|\varphi_2|$ replaced by 
$[\gamma]$, $|\gamma_1|$, $|\gamma_2|$
and $\pi$ replaced by $\zeta$.
From \Eref{Q_nP1_k_sum_done_purple1},
\begin{subequations}
\elabel{Q_nP1_k_sum_done_purple123_all}
\begin{align}
-
 \frac{\Aa(p_{n,\mtilde}^{(0)})}{2p_{n,\mtilde}^{(0)}}
 [\gamma]_{n,\mtilde}^{(0)}
&&& 
\mapsto 
\begin{aligned}[t]
&\zetatilde_{n+1,\mtilde+1}
&\text{for }\mtilde=1,\ldots,\Mtilde_n
\end{aligned}\\
 -
 \frac{\Ba(p_{n,\mtilde}^{(0)})}{2p_{n,\mtilde}^{(0)}}
 [\gamma]_{n,\mtilde}^{(0)} 
&&& 
\mapsto 
\begin{aligned}[t]
&\zetatilde_{n+1,\mtilde-1}
&\text{for }\mtilde=2,\ldots,\Mtilde_n 
\end{aligned}\\
 -
 \frac{\Aa(p^{(1)}_{n,m})}{2p^{(1)}_{n,m}}
 [\gamma]_{n,m}^{(1)}
&&& 
\mapsto 
\begin{aligned}[t]
&\zeta^{(1)}_{n+1,m+1}
&\text{for }m=-M^{(1)}_n,\ldots,M^{(1)}_n
\end{aligned}\\
 -
 \frac{\Ba(p^{(1)}_{n,m})}{2p^{(1)}_{n,m}}
 [\gamma]_{n,m}^{(1)}
&&& 
\mapsto 
\begin{aligned}[t]
&\zeta^{(1)}_{n+1,m-1}
&\text{for }m=-M^{(1)}_n,\ldots,M^{(1)}_n
\end{aligned}\\
 -
 \frac{\Aa(p^{(2)}_{n,m})}{2p^{(2)}_{n,m}}
 [\gamma]_{n,m}^{(2)}
&&& 
\mapsto
\begin{aligned}[t]
& \zeta^{(2)}_{n+1,m+1}
&\text{for }m=-M^{(2)}_n,\ldots,M^{(2)}_n
\end{aligned}\\
 -
 \frac{\Ba(p^{(2)}_{n,m})}{2p^{(2)}_{n,m}}
 [\gamma]_{n,m}^{(2)}
&&& 
\mapsto 
\begin{aligned}[t]
&\zeta^{(2)}_{n+1,m-1}
&\text{for }m=-M^{(2)}_n,\ldots,M^{(2)}_n
\end{aligned}\\
\frac{1}{L}
 \sum_{m=1}^{M_n} 
 \frac{1}{(p_{n,m})^2}
 [\gamma]_{n,m}
&&& 
\mapsto \zetatilde_{n+1,1} \ .
\end{align}
\end{subequations}

From \Eref{Q_nP1_k_sum_done_purple2}, 
\begin{subequations}
\begin{align}
 \frac{\Aa(a_1)}{2a_1}
 |\gamma_1|_n
&&& 
\mapsto \zeta^{(1)}_{n+1,1}\\
 \frac{\Ba(a_1)}{2a_1}
 |\gamma_1|_n
&&& 
\mapsto \zeta^{(1)}_{n+1,-1}\\
 -\frac{1}{La_1^2}
 |\gamma_1|_n
&&& 
\mapsto \zetatilde_{n+1,1} \ .
\end{align}
\end{subequations}

From \Eref{Q_nP1_k_sum_done_purple3}, 
\begin{subequations}
\begin{align}
 \frac{\Aa(a_2)}{2a_2}
 |\gamma_2|_n
&&& 
\mapsto \zeta^{(2)}_{n+1,1}\\
 \frac{\Ba(a_2)}{2a_2}
 |\gamma_2|_n
&&& 
\mapsto \zeta^{(2)}_{n+1,-1}\\
 -\frac{1}{La_2^2}
 |\gamma_2|_n
&&& 
\mapsto \zetatilde_{n+1,1} \ .
\end{align}
\end{subequations}

\subsection{Contributions to $\rho_{n+1,m}$}
Contributions to $\rho_{n+1,m}$ derive from \Eref{R_nP1_k_sum_done}
and follow a different pattern compared to $\pi$ and $\zeta$.
From \Eref{R_nP1_k_sum_done_violet1},
\begin{subequations}
\elabel{R_nP1_k_sum_done_purple123_all}
\begin{align}
 \Half{\Ab(p_{n,\mtilde}^{(0)})} (p_{n,\mtilde}^{(0)}+\xi^{-1})
 [\eta]_{n,\mtilde}^{(0)}
&&& 
\mapsto 
\begin{aligned}[t]
&\rhotilde_{n+1,\mtilde+1}
&\text{for }\mtilde=1,\ldots,\Mtilde_n
\end{aligned}\\
 \Half{\Bb(p_{n,\mtilde}^{(0)})} (p_{n,\mtilde}^{(0)}-\xi^{-1})
 [\eta]_{n,\mtilde}^{(0)}
&&& 
\mapsto 
\begin{aligned}[t]
&\rhotilde_{n+1,\mtilde-1}
&\text{for }\mtilde=2,\ldots,\Mtilde_n 
\end{aligned}\\
 \Half{\Ab(p^{(1)}_{n,m})} (p^{(1)}_{n,m}+\xi^{-1})
 [\eta]_{n,m}^{(1)}
&&& 
\mapsto 
\begin{aligned}[t]
&\rho^{(1)}_{n+1,m+1}
&\text{for }m=-M^{(1)}_n,\ldots,M^{(1)}_n
\end{aligned}\\
 \Half{\Bb(p^{(1)}_{n,m})} (p^{(1)}_{n,m}-\xi^{-1})
 [\eta]_{n,m}^{(1)}
&&& 
\mapsto 
\begin{aligned}[t]
&\rho^{(1)}_{n+1,m-1}
&\text{for }m=-M^{(1)}_n,\ldots,M^{(1)}_n
\end{aligned}\\
 \Half{\Ab(p^{(2)}_{n,m})} (p^{(2)}_{n,m}+\xi^{-1})
 [\eta]_{n,m}^{(2)}
&&& 
\mapsto 
\begin{aligned}[t]
&\rho^{(2)}_{n+1,m+1}
&\text{for }m=-M^{(2)}_n,\ldots,M^{(2)}_n
\end{aligned}\\
 \Half{\Bb(p^{(2)}_{n,m})} (p^{(2)}_{n,m}-\xi^{-1})
 [\eta]_{n,m}^{(2)}
&&& 
\mapsto 
\begin{aligned}[t]
&\rho^{(2)}_{n+1,m-1}
&\text{for }m=-M^{(1)}_n,\ldots,M^{(1)}_n \ .
\end{aligned}
\end{align}
\end{subequations}

From \Eref{R_nP1_k_sum_done_violet2}, 
\begin{subequations}
\begin{align}
 - \Half{\Ab(a_1)}(a_1+\xi^{-1})
 |\eta_1|_n
&&& 
\mapsto \rho^{(1)}_{n+1,1}\\
 - \Half{\Bb(a_1)}(a_1-\xi^{-1})
 |\eta_1|_n
&&& 
\mapsto \rho^{(1)}_{n+1,-1}\ . 
\end{align}
\end{subequations}

From \Eref{R_nP1_k_sum_done_violet3}, 
\begin{subequations}
\begin{align}
 - \Half{\Ab(a_2)}(a_2+\xi^{-1})
 |\eta_2|_n
&&& 
\mapsto \rho^{(2)}_{n+1,1}\\
 - \Half{\Bb(a_2)}(a_2-\xi^{-1})
 |\eta_1|_n
&&& 
\mapsto \rho^{(2)}_{n+1,-1}\ . 
\end{align}
\end{subequations}

\section{Iteration of $P$, $Q$ and $R$}
\seclabel{explicit_iteration_PQR}
This section contains the explicit form of the
iterative relations in \Erefs{P_nP1_k_integral}, \eref{Q_nP1_k_integral} and \eref{R_nP1_k_integral}
with coefficients in \Erefs{P_coeffs}, \eref{Q_coeffs} and \eref{R_coeffs}, in dimensionless
form.
Defining the dimensionless wavenumber $\Lambda_j = k_j \xi = 2\pi j \xibar$, we have

\begin{subequations}
\elabel{explicit_iteration_PQR}
\begin{align}
P_{n+1} (\Lambda_j) \Lambda_j^2 =  &-\Lambda_j \xibar \nubar
    \sum_{
    \substack{i\in\Zset\\
    i\ne0}}
\frac{\Lambda_j-\Lambda_i}{(\Lambda_j-\Lambda_i)^2+1}
\nonumber\\&\times
\left\{
\frac{\Lambda_i^2(\Lambda_i^2+\gammabar(2+\Pe))+\gammabar^2}{(\Lambda_i^2+\gammabar)(\Lambda_i^2+\gammabar(2+\Pe))} P_n(\Lambda_i)
+\frac{\gammabar}{\Lambda_i^2+\gammabar(2+\Pe)} Q_n(\Lambda_i)
- \frac{\gammabar \sqrt{\Pe \gammabar}}{(\Lambda_i^2+\gammabar)(\Lambda_i^2+\gammabar(2+\Pe))} \xi R_n(\Lambda_i)
\right\} \\
Q_{n+1} (\Lambda_j) \Lambda_j^2 = & -\Lambda_j \xibar \nubar
    \sum_{
    \substack{i\in\Zset\\
    i\ne0}}
\frac{\Lambda_j-\Lambda_i}{(\Lambda_j-\Lambda_i)^2+1}
\nonumber\\&\times
\left\{
\frac{\gammabar}{\Lambda_i^2+\gammabar(2+\Pe)} P_n(\Lambda_i)
+\frac{\Lambda_i^2+\gammabar}{\Lambda_i^2+\gammabar(2+\Pe)} Q_n(\Lambda_i)
- \frac{ \sqrt{\Pe \gammabar}}{\Lambda_i^2+\gammabar(2+\Pe)} \xi R_n(\Lambda_i)
\right\} \\
\xi R_{n+1} (\Lambda_j) \imag \Lambda_j =  & -\Lambda_j \xibar \nubar
    \sum_{
    \substack{i\in\Zset\\
    i\ne0}}
\frac{(\Lambda_j-\Lambda_i)\imag \Lambda_i}{(\Lambda_j-\Lambda_i)^2+1}
\nonumber\\&\times
\left\{
\frac{\gammabar\sqrt{\Pe\gammabar}}{(\Lambda_i^2+\gammabar)(\Lambda_i^2+\gammabar(2+\Pe))} P_n(\Lambda_i)
+\frac{\sqrt{\Pe\gammabar}}{\Lambda_i^2+\gammabar(2+\Pe)} Q_n(\Lambda_i)
+\frac{\Lambda_i^2+2\gammabar}{(\Lambda_i^2+\gammabar)(\Lambda_i^2+\gammabar(2+\Pe))} \xi R_n(\Lambda_i)
\right\}
\end{align}
\end{subequations}
with base case in \Eref{def_PQR1},
\begin{subequations}
\begin{align}
P_1(\Lambda_j) = 
Q_1(\Lambda_j) = &
- \frac{\diff}{2L} \frac{\nubar\xibar}{\Lambda_j^2+1} \ ,\\
R_1(\Lambda_j) = 0 \ .
\end{align}
\end{subequations}
Writing the iterative evolution of $P$, $Q$ and $R$
in this form shows the role of each of them.
We find that $R_n$, the part of $\HeteroPot_n$ odd in $\Lambda_j$, is the one that
carries the clearest signature of activity: 
$R_n$ is generated through $\Pe>0$ at $n=2$ and its effect in $P_{n+1}$ and
$Q_{n+1}$ is proportional to $\sqrt{\Pe}$.
Meanwhile, the even part of the effective interaction vertices
$\HomoPot_n$ and $\HeteroPot_n$, $P_n$ and $Q_n$ respectively,
play a similar role in each other's evolution and even in that of $R_n$, \Eref{explicit_iteration_PQR}.
Starting at $R_1=0$, the numerical value of $\xi R_n$ is small compared to $P_n$ and $Q_n$.
Thus, $P_n$ and $Q_n$, which are identical to each other in the passive case, and alternate sign with $n$,
follow a similar pattern in the active case, albeit somewhat modified by a non-zero $\xi R_n$.
The contribution from $\xi R_n$ effectively suppresses the amplitude of $P_n$ and $Q_n$ for odd $n$,
where $P_n$ and $Q_n$ are negative, while leaving their amplitude essentially unchanged for even $n$.

\section{Compressibility factor}
\seclabel{long_S1}
The compressibility factor $S_1=\ave{\cos(k_1(x_1-x_2))}$ defined in \cite{letter} as the lowest Fourier mode of the
two-point correlation function, or the lowest non-zero mode of the structure factor 
$S_j=\ave{\exp{\imag k_j (x_1-x_2)}}$,
can be calculated analytically following our iterative scheme detailed above.
Defining the factor $f=\sqrt{2+\Pe} = \sqrt{2\mathsf{D}_{\text{eff}}/\diff}$, the compressibility factor is, up to second order in $\nubar = \nu/(\diff\xi)$,
\begin{align}
S_1 = & -  \nubar \xibar \frac{ 2 \left(\Lambda_1^2+\gammabar \right) \left(\Lambda_1^2+ 2\gammabar\right) +\Pe \gammabar \Lambda_1^2 }{\left( \Lambda_1^2+1 \right) \left(\Lambda_1^2+\gammabar \right) \left(\Lambda_1^2+\gammabar(2+\Pe) \right)} 
\nonumber\\
& + \nubar^2\xibar \frac{1}{2 (\Lambda_1^2+4)(\Lambda_1^2+\gammabar)(\Lambda_1^2+\gammabar f^2)(\Lambda_1^2+(1+\sqrt{\gammabar})^2)(\Lambda_1^2+(1+\sqrt{\gammabar}f)^2)} \nonumber\\
&\times\Bigg[
\frac{(\Lambda_1^2+\gammabar)(\Lambda_1^2+2\gammabar)(\Lambda_1^2+(1+\sqrt{\gammabar})^2)}{1+\sqrt{\gammabar}f}
\left( \Lambda_1^2 +1+6\gammabar + \frac{\sqrt{\gammabar}}{f}
\left(2\Lambda_1^2 + 2\gammabar f^2+8-f^2\right)\right)
\nonumber\\
&
- \left( k_1^4\xi^4+\gammabar(1+f^2)\Lambda_1^2 + 2\gammabar^2\right)
\Bigg(
(\Lambda_1^2+(1+\sqrt{\gammabar})^2)(\Lambda_1^2+(1+\sqrt{\gammabar}f)^2)
\left(-1+ \frac{\Pe \gammabar^2}{(1-\gammabar)(1-\gammabar f^2)}\right)
\nonumber\\
&
-\Pe \frac{\sqrt{\gammabar}(\Lambda_1^2+4)}{1+\Pe}
\left(
\frac{\Lambda_1^2+(1+\sqrt{\gammabar})^2}{f(1-\gammabar f^2)}
+
\frac{\Lambda_1^2+(1+\sqrt{\gammabar}f)^2}{1-\gammabar}
\right)
\Bigg) \nonumber\\
&
+\frac{\Pe \gammabar (\Lambda_1^2+ 2\gammabar)}{(1+\Pe)(1-\gammabar)(1-\gammabar f^2)}
\Bigg(
2(1+\Pe)(2\gammabar-1)(\Lambda_1^2+(1+\sqrt{\gammabar})^2)(\Lambda_1^2+(1+\sqrt{\gammabar}f)^2) \nonumber\\
&
+\Pe (1+\sqrt{\gammabar} f)(1-\gammabar)(\Lambda_1^2+4)(\Lambda_1^2+(1+\sqrt{\gammabar})^2)
+(1+\sqrt{\gammabar})(1-\gammabar f^2)(\Lambda_1^2+4)(\Lambda_1^2+(1+\sqrt{\gammabar}f)^2)
\Bigg)\Bigg]
\nonumber\\
& +\OC\left(\nubar^3\right) \ . 
\elabel{S1_long}
\end{align}

\end{widetext}

\bibliography{../PRLdraft/articles,../PRLdraft/books}

\end{document}

%% file: gp_macros.tex
\usepackage{amsmath}
\usepackage{amsfonts}
\usepackage{cases}
\usepackage{xspace}
\usepackage{psfrag}
\usepackage{color}
\usepackage{xcolor}

\newcommand{\Aa}{A}
\newcommand{\Ab}{A}
\newcommand{\Ba}{B}
\newcommand{\Bb}{B}

\newcommand{\ave}[1]{\mathchoice{\left\langle #1 \right\rangle}{\langle #1 \rangle}{\langle #1 \rangle}{\langle #1 \rangle}}

\newcommand{\imag}{\mathring{\imath}}

\newcommand{\plaind}{\mathrm{d}}
\newcommand{\dint}[1]{\mathchoice{\!\plaind#1\,}{\!\plaind#1\,}{\!\plaind#1\,}{\!\plaind#1\,}}
\newcommand{\dbar}{\plaind\mkern-7mu\mathchar'26}
\newcommand{\deltabar}{\delta\mkern-8mu\mathchar'26}
\newcommand{\dintbar}[1]{\mathchoice{\!\dbar#1\,}{\!\dbar#1\,}{\!\dbar#1\,}{\!\dbar#1\,}}

\usepackage{dsfont}

\newcommand{\gpset}[1]{\mathds{#1}}

\newcommand{\canetset}[1]{{\mathchoice {\hbox{$\sf\textstyle #1\kern-0.4em #1$}}
{\hbox{$\sf\textstyle #1\kern-0.4em #1$}}
{\hbox{$\sf\scriptstyle #1\kern-0.3em #1$}}
{\hbox{$\sf\scriptscriptstyle #1\kern-0.2em #1$}}}}

\newcommand{\Zset}{\gpset{Z}}

\def\nbZ{{\mathchoice {\hbox{$\sf\textstyle Z\kern-0.4em Z$}}
{\hbox{$\sf\textstyle Z\kern-0.4em Z$}}
{\hbox{$\sf\scriptstyle Z\kern-0.3em Z$}}
{\hbox{$\sf\scriptscriptstyle Z\kern-0.2em Z$}}}}

\newcommand{\gpvec}[1]{\mathbf{#1}}

\newcommand{\xvec}{\gpvec{x}}

\newcommand{\sigmavec}{\boldsymbol{\sigma}}

\newcommand{\MA}{\mathcal{A}}

\newcommand{\OC}{\mathcal{O}}

\newcommand{\Mtilde}{M^{(0)}}
\newcommand{\mtilde}{{m}}

\newcommand{\pitilde}{{\pi}^{(0)}}

\newcommand{\rhotilde}{{\rho}^{(0)}}

\newcommand{\zetatilde}{{\zeta}^{(0)}}

\newcommand{\phidagger}{\phi^{\dagger}}

\newcommand{\Half}[1]{\mathchoice{\frac{#1}{2}}{((#1)/2)}{\frac{#1}{2}}{((#1)/2)}}
\newcommand{\half}{\mathchoice{\frac{1}{2}}{(1/2)}{\frac{1}{2}}{(1/2)}}

\newcommand{\ExpNB}[1]{\operatorname{exp}}
\renewcommand{\exp}[1]{\mathchoice{\mathrm{e}^{#1}}{\operatorname{exp}\left(#1\right)}{\operatorname{exp}\left(#1\right)}{\operatorname{exp}\left(#1\right)}}

\newcommand{\elabel}[1]{\label{eq:#1}}
\newcommand{\eref}[1]{(\ref{eq:#1})}
\newcommand{\Eref}[1]{Eq.~(\ref{eq:#1})}
\newcommand{\Erefs}[1]{Eqs.~(\ref{eq:#1})}

\newcommand{\seclabel}[1]{\label{sec:#1}}
\newcommand{\sref}[1]{Sec.~\ref{sec:#1}}

\newcommand{\aref}[1]{App.~\ref{sec:#1}}

\newcommand{\flabel}[1]{\label{fig:#1}}
\newcommand{\fref}[1]{Fig.~\ref{fig:#1}}

\newcommand{\latin}[1]{{\it #1}}
\newcommand{\ie}{\latin{i.e.}\@\xspace}

\newcommand{\etc}{\latin{etc.}\@\xspace}

\newlength \standardfigwidth
\DeclareMathAlphabet{\matheub}{U}{eur}{m}{n}

\newcounter{exercise}
{\addtocounter{exercise}{1}\begin{center}\begin{minipage}{0.8\linewidth}\textbf{Exercise
\arabic{exercise}:}\begin{itshape}}
{\end{itshape}\end{minipage}\end{center}}

\makeatletter
\newcommand{\creat}[3][]{\@ifempty{#1}{#2^{\dagger}}{\left(#2^{\dagger}\right)^{#1}}\@ifempty{#3}{}{\!(#3)}}

\newcommand{\creatDoi}[3][]{\@ifempty{#1}{\tilde{#2}}{\left(\tilde{#2}\right)^{#1}}\@ifempty{#3}{}{(#3)}}

\newcommand{\annih}[3][]{#2\@ifempty{#1}{}{^{#1}}\@ifempty{#3}{}{(#3)}}

\makeatother

% ...

\newlength{\bibmarkkeyAleft}

\newlength{\bibmarkkeyBleft}

\newlength{\bibmarkkeyCleft}

\newlength{\bibmarkkeyDleft}

\usepackage{wasysym}

\makeatletter
\newcommand{\pushright}[1]{\ifmeasuring@#1\else\omit\hfill$\displaystyle#1$\fi\ignorespaces}
\newcommand{\pushleft}[1]{\ifmeasuring@#1\else\omit$\displaystyle#1$\hfill\fi\ignorespaces}
\makeatother

\newcommand{\diff}{\mathsf{D}_x}
\newcommand{\drift}{\mathsf{w}}
\newcommand{\HomoPot}{\Phi}
\newcommand{\HeteroPot}{\Psi}
\newcommand{\tHomoPot}{\widetilde{\Phi}}
\newcommand{\tHeteroPot}{\widetilde{\Psi}}
\newcommand{\pairPot}{W}
\newcommand{\pairPotInf}{\Upsilon}

%% file: macros_diagrams.tex
\usetikzlibrary{patterns}
\tikzset{
xxtsubstrate/.style={decorate, 
line width=1pt,
draw=cyan, 
decoration=snake, 
segment amplitude=0.75mm, 
line after snake=0.25mm,
line before snake=0.25mm
},
tsubstrate/.style={decorate, 
line width=1pt,
draw=cyan, 
decoration=snake, 
segment amplitude=0.5mm, 
segment length=5pt,
segment amplitude=0.2mm, 
line after snake=1mm,
line before snake=1mm
},
Bsubstrate/.style={decorate, 
line width=1pt,
draw=cyan, 
decoration=snake,
segment length=5pt,
segment aspect=0,
segment amplitude=0.5mm, 
line after snake=0mm,
line before snake=0mm
},
substrate/.style={decorate, 
line width=1pt,
draw=cyan, 
snake=snake, 
decoration=snake, 
segment length=5pt,
segment amplitude=0.5mm, 
line after snake=0.5mm,
line before snake=0.5mm
},
activity/.style={very thick,draw=red,postaction={decorate},
decoration={markings,mark=at position .5 with
{\arrow[draw=red]{>}}}},
cactivity/.style={very thick,draw=yellow,postaction={decorate},
decoration={markings,mark=at position .5 with
{\arrow[draw=yellow]{<}}}},
tactivity/.style={thick,draw=red,postaction={decorate},
decoration={markings,mark=at position .5 with
{\arrow[draw=red]{>}}}},
tEPSactivity/.style={thick,draw=red,postaction={decorate},
decoration={markings,mark=at position .55 with
{\arrow[draw=red]{>}}}},
tAactivity/.style={thick,draw=red},
Aactivity/.style={thick,draw=black},
Bactivity/.style={very thick,draw=black,dashed},
dotactivity/.style={very thick,draw=black,dotted},
Cactivity/.style={very thick,draw=black,decorate,decoration=snake},
Baractivity/.style={very thick,draw=black},
tSactivity/.style={thick,draw=red,postaction={decorate},
decoration={markings,mark=at position .7 with
{\arrow[draw=red]{>}}}},
Sactivity/.style={very thick,draw=red,postaction={decorate},
decoration={markings,mark=at position .7 with
{\arrow[draw=red]{>}}}}
}
\tikzset{
  terminal/.style = {
    rectangle,
    align = center,
    minimum size = 6mm,
    rounded corners = 3mm,
    very thick,
    draw = black!50,
    top color = white,
    bottom color = black!20,
  }
}

\tikzset{
tAsubstrate/.style={decorate, 
thick,
decoration=snake, 
segment amplitude=0.5mm, 
segment length=5pt,
segment amplitude=0.2mm, 
line after snake=1mm,
line before snake=1mm
},
DasHsubstrate/.style={draw=none,
postaction={decorate,decoration={markings,
        mark=at position 0.25 with {\draw[cyan] (0,-1.2mm) -- (0,1.2mm);}
        }}
},
tactivity/.style={thick,draw=red,postaction={decorate},
decoration={markings,mark=at position .5 with
{\arrow[draw=red]{>}}}},
tAactivity/.style={thick,draw=red},
tAactivityDasheD/.style={thick,draw=red,postaction={decorate,decoration={markings,
        mark=at position 0.75 with {\draw[red] (0,-1.2mm) -- (0,1.2mm);}
        }}},
DasheDtAactivity/.style={thick,draw=red,postaction={decorate,decoration={markings,
        mark=at position 0.25 with {\draw[red] (0,-1.2mm) -- (0,1.2mm);}
        }}},
tAactivityDA/.style={thick,draw=red,postaction={decorate,decoration={markings,
        mark=at position 0.6 with {\draw[red] (0,-1.2mm) -- (0,1.2mm);}
        }}},
AactivityDA/.style={thick,draw=black,postaction={decorate,decoration={markings,
        mark=at position 0.6 with {\draw[black,thin] (0,-1.2mm) -- (0,1.2mm);}
        }}},
DAtAactivity/.style={thick,postaction={decorate,decoration={markings,
        mark=at position 0.4 with {\draw (0,-1.2mm) -- (0,1.2mm);}
        }}},
potential/.style={thick,draw=black,dashed},
potentialDasheD/.style={thick, draw=black,densely dashed,postaction={decorate,decoration={markings,
        mark=at position 0.4 with {\draw[black,solid,thin] (0,-1.1mm) -- (0,1.1mm);}
        }}},
DasheDpotential/.style={thick,draw=black,densely dashed,postaction={decorate,decoration={markings,
        mark=at position 0.25 with {\draw[solid,thin] (0,-1.2mm) -- (0,1.2mm);}
        }}},
}

\tikzset{half paths/.style 2 args={%
  decoration={show path construction,
    lineto code={
      \draw [#1,out=150,in=30] (\tikzinputsegmentfirst) to[bend right]
         ($(\tikzinputsegmentfirst)!0.5!(\tikzinputsegmentlast)$);
      \draw [#2,out=150,in=30] ($(\tikzinputsegmentfirst)!0.5!(\tikzinputsegmentlast)$)
        to[bend right] (\tikzinputsegmentlast);
    }
  }, decorate
}}

\newcommand{\LoopDiagramfi}[9]{
    \tikz[baseline=-2.5pt]{
    \begin{scope}[xshift=-1.4cm]
    \draw[#1] (-0.4,0.4) -- (0,0.3);
    \draw[draw=none,in=30,out=150]  (-0.4,0.4) -- node[pos=0.4,above=0.1cm] {} 
    node[pos=0.4,sloped] {\tikz{\draw[#2] (0,-1.2mm) -- (0,1.2mm);}} (0,0.3);
    \draw[DasheDpotential] (0,0.3) -- (0,-0.3);
    \draw[DasheDpotential] (1,0.3) -- +(0,-0.6);
    \draw[#6] (-0.4,-0.4) -- (0,-0.3);
    \end{scope}
    \draw[#3,out=150,in=180] (-0.4,0.3) to[bend right] (-0.9,0.55);
    \draw[#1,out=180,in=30] (-0.9,0.55) to[bend right] (-1.4cm,0.3);
    \draw[draw=none,in=180,out=150]  (-0.4,0.3) to[bend right] node[pos=0.4,above=0.1cm] {$#5$} 
    node[pos=0.4,sloped] {\tikz{\draw[#4] (0,-1.2mm) -- (0,1.2mm);}} (-1,0.55);
    \draw[#7,in=-180,out=210] (-0.4,-0.3) to[bend left] (-0.9cm,-0.55);
    \draw[#6,out=180,in=-30] (-0.9cm,-0.55) to[bend left] (-1.4cm,-0.3);
    \draw[draw=none,in=-30,out=210]  (-0.4,-0.3) to[bend left] node[pos=0.4,below=0.1cm] {$#9$} 
    node[pos=0.4,sloped] {\tikz{\draw[#8, thick] (0,-1.2mm) -- (0,1.2mm);}} (-1,-0.55);
    }
}

\newcommand{\LoopDiagramreverse}[9]{
    \tikz[baseline=-2.5pt]{
    \begin{scope}[xshift=-1.4cm]
    \draw[#1] (-0.4,0.4) -- (0,0.3);
    \draw[draw=none,in=30,out=150]  (-0.4,0.4) -- node[pos=0.4,above=0.1cm] {} 
    node[pos=0.4,sloped] {\tikz{\draw[#2, thick] (0,-1.2mm) -- (0,1.2mm);}} (0,0.3);
    \draw[DasheDpotential] (0,0.3) -- (0,-0.3);
    \draw[DasheDpotential] (1,-0.3) -- +(0,0.6);
    \draw[#6] (-0.4,-0.4) -- (0,-0.3);
    \end{scope}
    \draw[#3,out=150,in=180] (-0.4,0.3) to[bend right] (-0.9,0.55);
    \draw[#1,out=180,in=30] (-0.9,0.55) to[bend right] (-1.4cm,0.3);
    \draw[draw=none,in=180,out=150]  (-0.4,0.3) to[bend right] node[pos=0.4,above=0.1cm] {$#5$} 
    node[pos=0.4,sloped] {\tikz{\draw[#4, thick] (0,-1.2mm) -- (0,1.2mm);}} (-1,0.55);
    \draw[#7,in=-180,out=210] (-0.4,-0.3) to[bend left] (-0.9cm,-0.55);
    \draw[#6,out=180,in=-30] (-0.9cm,-0.55) to[bend left] (-1.4cm,-0.3);
    \draw[draw=none,in=-30,out=210]  (-0.4,-0.3) to[bend left] node[pos=0.4,below=0.1cm] {$#9$} 
    node[pos=0.4,sloped] {\tikz{\draw[#8, thick] (0,-1.2mm) -- (0,1.2mm);}} (-1,-0.55);
    }
}

\newcommand{\LoopDiagramPot}{\HomoPot_n(k')}

\newcommand{\LoopDiagram}[9]{
    \tikz[baseline=-2.5pt]{
    \begin{scope}[xshift=-1.4cm]
    \draw[#1] (-0.4,0.4) -- (0,0.3);
    \draw[draw=none,in=30,out=150]  (-0.4,0.4) -- node[pos=0.4,above=0.1cm] {} 
    node[pos=0.4,sloped] {\tikz{\draw[#2, thick] (0,-1.2mm) -- (0,1.2mm);}} (0,0.3);
    \draw[DasheDpotential] (0,0.3) -- (0,-0.3);
    \draw[#6] (-0.4,-0.4) -- (0,-0.3);
    \end{scope}
    \draw[pattern=north west lines] (0,0) circle (0.7cm);
    \draw[fill=white] (0,0) node[circle,fill=white,inner sep=1pt,opacity=0.9] {\footnotesize $\LoopDiagramPot$};
    \draw[#3,out=150,in=180] (150:0.7cm) to[bend right] (-1,0.501);
    \draw[#1,out=180,in=30] (-1,0.501) to[bend right] (-1.4cm,0.3);
    \draw[draw=none,in=180,out=150]  (150:0.7cm) to[bend right] node[pos=0.4,above=0.1cm] {$#5$} 
    node[pos=0.4,sloped] {\tikz{\draw[#4, thick] (0,-1.2mm) -- (0,1.2mm);}} (-1,0.501);
    \draw[#7,in=-180,out=210] (210:0.7cm) to[bend left] (-1cm,-0.501);
    \draw[#6,out=180,in=-30] (-1cm,-0.501) to[bend left] (-1.4cm,-0.3);
    \draw[draw=none,in=-30,out=210]  (210:0.7cm) to[bend left] node[pos=0.4,below=0.1cm] {$#9$} 
    node[pos=0.4,sloped] {\tikz{\draw[#8, thick] (0,-1.2mm) -- (0,1.2mm);}} (-1,-0.501);
    }
}

\tikzset{
    position label/.style={
       below = 3pt,
       text height = 1.5ex,
       text depth = 1ex
    },
   brace/.style={
     decoration={brace, mirror},
     decorate
   }
}